\documentclass[preprint,amsmath,amssymb,aps]{revtex4-1}
\usepackage{graphicx}
\usepackage{dcolumn}
\usepackage{bm}
\usepackage{float}
\usepackage{color}
\usepackage{amssymb}
\usepackage{tikz}
\usetikzlibrary{shapes}
\usepackage{xcolor}

\definecolor{verdecito}{RGB}{0,168,89}
\definecolor{rojito}{RGB}{237,50,55}
\definecolor{azulito}{RGB}{0,0,255}
\definecolor{violetita}{RGB}{111,0,146}
\definecolor{orangered}{RGB}{255,69,0}
\definecolor{sky}{RGB}{0,191,255}
\definecolor{orangeredII}{RGB}{205,55,0}
\definecolor{verdecito}{RGB}{0,168,89}
\definecolor{sepia}{RGB}{94,38,18}

\newcommand{\verdcirc}{\raisebox{0.5pt}{\tikz{\node[draw,scale=0.5,circle,verdecito,fill=verdecito](){};}}}

\newcommand{\magcirc}{\raisebox{0.5pt}{\tikz{\node[draw,scale=0.5,circle,magenta](){};}}}
\newcommand{\skycirc}{\raisebox{0.5pt}{\tikz{\node[draw,scale=0.3,circle,sky,fill=sky](){};}}}
\newcommand{\violdiamond}{\raisebox{0.5pt}{\tikz{\node[draw,scale=0.3,diamond,violetita,fill=violetita](){};}}}
\newcommand{\bckdotted}{\raisebox{2pt}{\tikz{\draw[black,dotted,line width=1.5pt](0,0) -- (5mm,0);}}}
\newcommand{\bluedotted}{\raisebox{2pt}{\tikz{\draw[blue,dotted,line width=1.5pt](0,0) -- (5mm,0);}}}
\newcommand{\magdotted}{\raisebox{2pt}{\tikz{\draw[magenta,dotted,line width=1pt](0,0) -- (5mm,0);}}}
\newcommand{\greendotted}{\raisebox{2pt}{\tikz{\draw[green,dotted,line width=2.3pt](0,0) -- (5mm,0);}}}
\newcommand{\blackdotted}{\raisebox{2pt}{\tikz{\draw[black,dotted,line width=2.3pt](0,0) -- (5mm,0);}}}
\newcommand{\reddashed}{\raisebox{2pt}{\tikz{\draw[red,dashed,line width=1pt](0,0) -- (5mm,0);}}}
\newcommand{\magdashed}{\raisebox{2pt}{\tikz{\draw[magenta,dashed,line width=1pt](0,0) -- (5mm,0);}}}

\newcommand{\verdecitodashdot}{\raisebox{2pt}{\tikz{\draw[verdecito,dashdotted,line width=1pt](0,0) -- (5mm,0);}}}
\newcommand{\bluesolid}{\raisebox{2pt}{\tikz{\draw[blue,solid,line width=1pt](0,0) -- (5mm,0);}}}

\newcommand{\blackdashdotted}{\raisebox{2pt}{\tikz{\draw[black,dashdotted,line width=1pt](0,0) -- (5mm,0);}}}
\newcommand{\redsolid}{\raisebox{2pt}{\tikz{\draw[red,solid,line width=1pt](0,0) -- (5mm,0);}}}
\newcommand{\grsolid}{\raisebox{2pt}{\tikz{\draw[green,solid,line width=1pt](0,0) -- (5mm,0);}}}
\newcommand{\verdecitosolid}{\raisebox{2pt}{\tikz{\draw[verdecito,solid,line width=1pt](0,0) -- (5mm,0);}}}
\newcommand{\verdash}{\raisebox{2pt}{\tikz{\draw[verdecito,dashed,line width=1pt](0,0) -- (5mm,0);}}}
\newcommand{\bluedashdotted}{\raisebox{2pt}{\tikz{\draw[blue,dashdotted,line width=1pt](0,0) -- (5mm,0);}}}
\newcommand{\blacksolid}{\raisebox{2pt}{\tikz{\draw[black,solid,line width=1pt](0,0) -- (5mm,0);}}}

\newcommand{\rojsq}{\raisebox{0.5pt}{\tikz{\node[draw,scale=0.5,regular polygon, regular polygon sides=4,rojito,fill=rojito](){};}}}
\newcommand{\blcktrg}{\raisebox{0.5pt}{\tikz{\node[draw,scale=0.5,regular polygon, regular polygon sides=3,fill=black](){};}}}
\newcommand{\azptg}{\raisebox{0.5pt}{\tikz{\node[draw,scale=0.5,regular polygon, regular polygon sides=5,azulito,fill=azulito](){};}}}

\newcommand{\rojsqII}{\raisebox{0.5pt}{\tikz{\node[draw,scale=0.3,regular polygon, regular polygon sides=4,rojito](){};}}}
\newcommand{\blcktrgII}{\raisebox{0.5pt}{\tikz{\node[draw,scale=0.3,regular polygon, regular polygon sides=3](){};}}}
\newcommand{\azcircII}{\raisebox{0.5pt}{\tikz{\node[draw,scale=0.5,circle,azulito](){};}}}
\newcommand{\magtrg}{\raisebox{0.5pt}{\tikz{\node[draw,scale=0.3,regular polygon, regular polygon sides=3,magenta,fill=magenta](){};}}}
\newcommand{\orgIItrginv}{\raisebox{0.5pt}{\tikz{\node[draw,rotate=180,scale=0.3,regular polygon, regular polygon sides=3,orangeredII,fill=orangeredII](){};}}}
\newcommand{\sepiacross}{\raisebox{0.5pt}{\tikz{\node[draw,scale=0.4,cross out,sepia,line width=1.5pt](){};}}}

\newcommand{\blackcircle}{\raisebox{0.5pt}{\tikz{\node[draw,scale=0.5,circle,black,line width=0.6pt](){};}}}
\newcommand{\redsquare}{\raisebox{0.5pt}{\tikz{\node[draw,scale=0.3,regular polygon, regular polygon sides=4,red](){};}}}
\newcommand{\bluetriangle}{\raisebox{0.5pt}{\tikz{\node[draw,scale=0.3,regular polygon, regular polygon sides=3,blue,fill=blue](){};}}}

\newcommand{\skydiamond}{\raisebox{0.5pt}{\tikz{\node[draw,scale=0.3,diamond,sky](){};}}}
\newcommand{\magsolid}{\raisebox{2pt}{\tikz{\draw[magenta,solid,line width=1pt](0,0) -- (5mm,0);}}}

\newcommand{\blackdashed}{\raisebox{2pt}{\tikz{\draw[black,dashed,line width=1pt](0,0) -- (5mm,0);}}}
\newcommand{\verdecitodashdotted}{\raisebox{2pt}{\tikz{\draw[verdecito,dashdotted,line width=1pt](0,0) -- (5mm,0);}}}
\newcommand{\skydiamondII}{\raisebox{0.5pt}{\tikz{\node[draw,scale=0.3,diamond,sky,fill=sky](){};}}}

\begin{document}

\title{Multiple outbreaks in epidemic spreading with local vaccination and limited vaccines}

\author{Mat\'ias, A, {Di Muro}}
 \affiliation{Instituto de Investigaciones F\'isicas de Mar del Plata
  (IFIMAR)-Departamento de F\'isica, Facultad de Ciencias Exactas y
  Naturales, Universidad Nacional de Mar del Plata-CONICET, Funes
  3350, (7600) Mar del Plata, Argentina.}
  \email{mdimuro@mdp.edu.ar}
\author{Lucila G. {Alvarez-Zuzek}}%
\affiliation{Instituto de Investigaciones F\'isicas de Mar del Plata
  (IFIMAR)-Departamento de F\'isica, Facultad de Ciencias Exactas y
  Naturales, Universidad Nacional de Mar del Plata-CONICET, Funes
  3350, (7600) Mar del Plata, Argentina.}
\author{Shlomo Havlin}
 \affiliation{Department of Physics, Bar-Ilan University, Ramat-Gan 52900, Israel.}
  \author{Lidia A. Braunstein}
  \affiliation{Instituto de Investigaciones F\'isicas de Mar del Plata
  (IFIMAR)-Departamento de F\'isica, Facultad de Ciencias Exactas y
  Naturales, Universidad Nacional de Mar del Plata-CONICET, Funes
  3350, (7600) Mar del Plata, Argentina.}
  \affiliation{Center for Polymer
  Studies, Boston University, Boston, Massachusetts 02215, USA}


\begin{abstract} 
How to prevent the spread of human diseases is a great challenge for the scientific community and so far there are many studies in which immunization strategies have been developed. However, these kind of strategies usually do not consider that medical institutes may have limited vaccine resources available. In this manuscript, we explore the Susceptible-Infected-Recovered (SIR) model with local dynamic vaccination, and considering limited vaccines. In this
model, susceptibles in contact with an infected individual, are vaccinated -with probability $\omega$- and then get infected -with probability $\beta$. However, when the fraction of immunized individuals reaches a threshold $V_L$, the vaccination stops, after which only the infection is possible. In the steady state, besides the critical points $\beta_c$ and $\omega_c$ that separate a non-epidemic from an epidemic phase, we find for a range of $V_L$ another transition points, $\beta^*>\beta_c$ and $\omega^*<\omega_c$, which correspond to a novel discontinuous phase transition. This critical value separates a phase where the amount of vaccines is sufficient, from a phase where the disease is strong enough to exhaust all the vaccination units. For a disease with fixed $\beta$, the vaccination probability $\omega$ can be controlled in order to drastically reduce the number of infected individuals, using efficiently the available vaccines.

Furthermore, the temporal evolution of the system close to $\beta^*$ or $\omega^*$, shows that after a peak of infection the system enters into a quasi-stationary state, with only a few infected cases. But if there are no more vaccines, these few infected individuals could originate a second outbreak, represented by a second peak of infection. This state of apparent calm, could be dangerous since it may lead to misleading conclusions and to an abandon of the strategies to control the disease.
\end{abstract}

\maketitle
\section{Introduction}

Human interactions have a structure that can be well described in the
form of a complex network \cite{Boc_01,barrat_04,New_10,Coh_10}. In the
last few years, new technologies allowed us to record large amount of
data of contact patterns
\cite{Catt_01,gonzalez_08,gardenes_08,volz_11}. This data has become
accessible to researches that use data-driven network modeling approaches to
analyze and understand spreading in social systems, for example, how epidemic and even rumors spread in real populations.

Scientists have focused
\cite{Bailey_75,Ander_91,past_03,Arenas_16,castellano_10},
on modeling and analyzing disease spreading since it can lead to
catastrophic health consequences as well as large economic
losses. Several mathematical approaches have been developed and used to study
different epidemic models, improving the understanding of
disease spreading on complex networks \cite{Pastor_15,Braunstein_16} (and references
therein).

Since one of the goals of health authorities is to minimize health
catastrophes and economic impact of health policies, many studies have
focused on establishing immunization and mitigation strategies for
enhancing the functionality of a society and reduce the economic cost
\cite{Coh_03,Gallos_07}. For example, vaccination programs
\cite{wang_16} are very efficient in providing immunity to individuals
and as a consequence, the final number of infected people decreases
considerably. However, these strategies are usually very expensive and
unrealistic, because vaccines against new strains are usually not
available during the initial propagation stage. As a consequence,
non-pharmaceutical interventions are needed to protect the
society. One of the most effective and studied strategy to slow an
epidemic is quarantine. However, it has the disadvantage that full
isolation has a negative impact on the economy of the region and it is
difficult to implement it in a large population. Thus, it is important
to find a balance between these two strategies. Another policy, such
as social distancing strategies, have been modeled and implemented in
order to reduce the average contact time between individuals
\cite{Valdez_13,wang_12}. This kind of strategies, usually include
closing schools, cough etiquette, travel restrictions, intermittent
connections, etc. Unfortunately, in most cases, these strategies do
not prevent a pandemic, but only delays its spread.

One of the most remarkable cases of a disease spreading was the
pandemic occurred by the H1N1 strain in 2009, which caused about
$15.000$ deaths. Initially, the disease propagated over the network of
close contacts, and then through the airline network, transporting
infected individuals to different cities, thus spreading the disease all over the world. One of the most used models to mimic these kind of
epidemics is the Susceptible-Infected-Recovered ($SIR$) model
\cite{New_05,moreno_02,volzsir,miller_11}. In this model an individual
can be in only one of three possible states: Susceptible (S), Infected
(I) or Recovered (R). An individual in state S in contact with an I,
changes to an I state with probability $\beta$. After a period of time
$t_r$ the infected individual changes to an $R$ state and stops
transmitting the disease. This model presents, in the steady state,
two regimes governed by an effective probability of contagion
$T=T_{\beta,t_r}=1-(1-\beta)^{t_r}$, such that for $ T \leq T_c $ the
system is in an epidemic-free phase and for $ T> T_c $ it is in an
epidemic phase, where the disease reaches a high fraction of the
population. The $SIR$ model has been also successfully applied to
model the case of SARS and other diseases of influenza type
\cite{Colizza_06,meyers_sars}.  For decades, researchers have studied
different scenarios of the $SIR$ epidemic model
\cite{Val_12,volz_07,xia_12,Buono_14} and develop mitigation
strategies to prevent the epidemic
\cite{Buono_15,Alvarez-zuzek_15,Lag_01, alvarez2015,wang_16}, such as
isolation, quarantine and random and targeted
vaccination. Particularly, Valdez {\it et al.}  \cite{valdez2012}
studied, using the {\it SIR} model, the effect of an intermittent
social distancing strategy on the propagation of epidemics in adaptive
complex networks. Based on local information, a susceptible individual
interrupts the contact with an infected individual with a certain
probability and restores it after a fixed period of time. In a similar
way, Ref. \cite{Valdez_Ebola} extended the model and was able to
successfully predict the date of extinction of the Ebola's outbreak in
Liberia in $2014$. Ebola outbreaks have been studied by the scientific
community due to the high impact of this epidemic on certain regions
of southwest Africa, mainly in Guinea, Sierra Leona and
Liberia. Fortunately, there exists available data for the scientific
community enabling to study more accurately the behavior and
propagation of this disease \cite{Valdez_Ebola,Gomes_14}.

During this Ebola outbreak, a vaccine trial has been performed and used in
Guinea in 2015 in the capital city. It has been found that the
strategy of the vaccine trial applied to mitigate the transmission of
Ebola-Virus-Disease ($EVD$) in Guinea in 2015, was very efficient
\cite{lancet_ebola}. The vaccine trial tested the efficacy of an
experimental vaccine against Ebola. The trial used a ``ring''
vaccination’ strategy based on the approach that was used to eradicate
the smallpox \cite{Smallpox}. This involves the identification of a
newly diagnosed Ebola case, and then the vaccination of all his contacts and
the contacts of those contacts, which are usually their family
members, neighbors, co-workers and friends. In practical terms, the
close contacts of a newly identified Ebola case have been vaccinated,
if they consent to it. This is the basis of the motivation of the
present study where we immunize the neighbors of an infected
individual in network models.
Moreover, the amount of vaccines, sometime due to economic
restrictions, may not be enough to protect all the susceptible
neighbors of the infected individuals during the whole spreading
process. 
Thus, we propose and study here models with the scenario of limited
vaccines availability, i.e. not enough for all the vulnerable
population.  Generally speaking, the available resources to control,
avoid, or maybe enhance a spreading process are limited, and many
studies have focused in how to optimally use these scarce resources
against a disease \cite{Lokhov}, or even to deal with illicit drug
usage \cite{Drug}.

We
are interested in testing how this limited amount of accessible vaccines
affects the spread of the epidemic.  Motivated by this, we present
here a model of localized vaccination that mimics the vaccine trial in
Guinea, in which only neighbors of infected individuals could be
immunized using limited vaccination units. This could help to
understand how the propagation of a disease is affected by the
localized and limited vaccination.
Therefore, the question we wish to answer
is how to use, in an efficiently way, this limited amount of vaccines
with the aim of reducing the propagation of an epidemic.

In our model,
the state of an individual can be
Susceptible (S), Vaccinated (V), Infected (I) or Recovered (R). A
susceptible individual in contact with an infected one will become
vaccinated with probability $\omega$, until the total number of vaccines
$V_L$ is used. If he does not become vaccinated, then with probability
$\beta$ he will become infected. Then, after
a certain period of time $t_r$, the infected individuals will recover.

Using an edge-based compartmental model and the generating functions
theory \cite{miller_11,miller_12}, we obtain and study the evolution
equations for the fraction of S, I, R and V individuals and find a
perfect agreement between theoretical and simulation results.
Then, we study the steady state of the epidemic process, for which
there is no more infected individuals. We find two different phases,
an epidemic and a non-epidemic phase, separated by a critical
threshold $\beta_c$, which depend on $\omega$. Below $\beta_c$ the
disease can not spread and the fraction of recovered individuals $R$
approaches to zero. On the other hand if $\beta$ is fixed, there is a
critical vaccination probability $\omega_c$, above which the epidemic
can not develop, since the disease is successfully blocked by immunized
individuals.

Above $\beta_c$, depending on the parameters, we find either a
continuous phase transition or a discontinuous phase transition for a second epidemic break at higher critical threshold $\beta^*$, which
depends on $V_L$. Similarly for fixed $\beta$, below $\omega_c$ we
find a discontinuous transition for $\omega=\omega^\ast$ in the
fraction of recovered individuals. In Refs. \cite{Bottcher_15,Chen2,Chen_18}
the authors also found a discontinuous transition but in the density
of infected individuals, and using an endemic epidemic model
(susceptible-infected-susceptible), where the recovery of sick
individuals depends on the availability of healing resources.

We also find, that depending on the parameters there are values of
$\beta=\beta^\dag>\beta^\ast$ and $\omega=\omega^\dag<\omega^\ast$
that characterize a crossover between two regimes, one for which the
available amount of vaccines is sufficient to immunize the population
during the whole process and the other in which it is not.

\section{The Model}

\begin{figure}
  \begin{center}
    \includegraphics[width=0.65\textwidth]{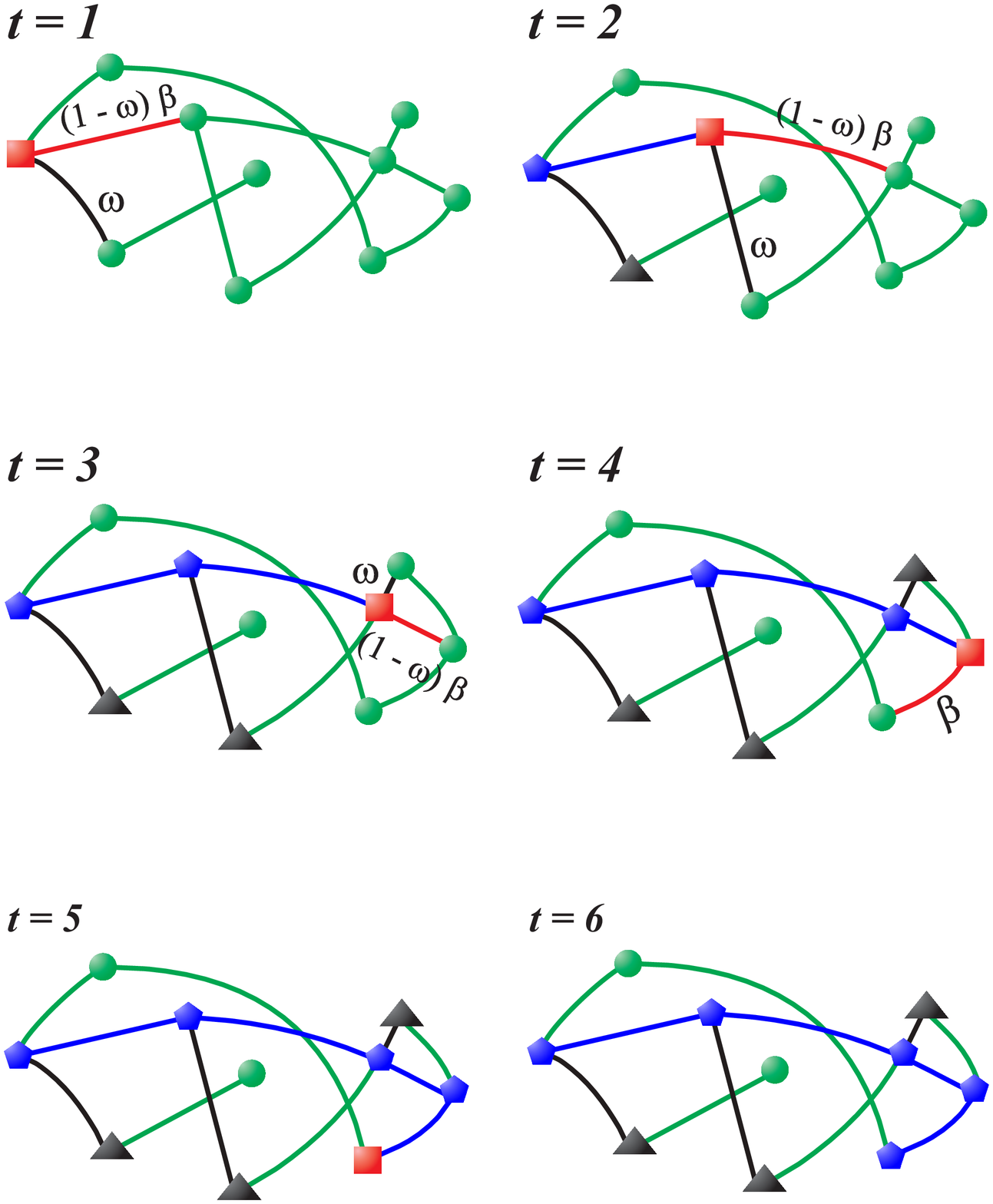}
  \end{center}
  \caption{Schematic demonstration of the rules of the model with limited vaccines
    for a small network of size $N=10$, a recovery time $t_r=1$, and a vaccination limit of
    $V_L=3/10$. The color of the nodes represents the different states:
    susceptible (S) (\protect\verdcirc), infected (I)
    (\protect\rojsq), vaccinated (V) (\protect\blcktrg), recovered (R) (\protect\azptg). At $t=1$ the patient zero induces
    the vaccination on its neighbors with probability $\omega$ or infects
    them with probability $(1-w)\beta$. After a time $t_r$ the
    infected individuals move to the recovered state R, in this case
    $t_r=1$. At $t=4$ all the vaccines were used and hence the
    infected individual only tries to infect susceptible
      neighbors with probability $\beta$. At $t=6$ the steady state
    is reached and nodes can only be in state S, R or
    V.}\label{scheme}
\end{figure}

At the initial state, all the individuals in the network are
susceptible except one, the patient zero, which is in the infected
compartment or state. Before spreading the disease, all the
susceptible neighbors of this individual receive a vaccine with
probability $\omega$. Notice that this vaccination is local and
dynamic, and is done through the links that connect infected to
susceptible nodes. Then the patient zero will try to infect all its
neighbors that have not been vaccinated, and this event will occur
with probability $(1-\omega)\beta$. In the next time step of the
process, all the susceptible neighbors of the infected nodes will be
vaccinated with probability $\omega$ again or will be infected with
probability $(1-\omega)\beta$. The infected individuals will move to
the recovered state R after $t_r$ units of time since they become infected, and the vaccinated or immunized individuals will remain in
state V. While the disease spreads through the population, the number
of vaccinated people increases, until the health institutes run out of
vaccines. We define $V_L$ as the fraction of available vaccines over
the entire population. When this limit is reached, no more individuals
can be immunized and hence those in the infected state will infect
their neighbors with probability $\beta$. In the steady state the
epidemic is over when the fraction of infected individuals is zero,
thus the individuals can only be in state S, R or V. For demonstration
of the model see Fig. \ref{scheme}.

This model is clearly different from random vaccination, in which a
fraction of individuals selected at random are immunized. In the
random strategy, some vaccinated individuals may never be in contact
with an infected individual and thus would not be vaccinated, in the present strategy. On the
other hand the dynamical vaccination intends to create a barrier of
immunized individuals that could stop the spreading of the disease, thus
making a more effective usage of the available vaccines.

\section{Theoretical Formalism}

The edge-based compartmental model (EBCM)
\cite{miller_11,volz_11}, was applied to model the SIR and was adapted by Valdez {\it et
  al.} for discrete time, and a fixed recovery time $t_r$ \cite{Val_12}. We can solve
theoretically the evolution and the steady state of this model with
unlimited vaccines \cite{Alvarez_2018} and also adapt it here for the
limited case.  The EBCM is based on a generating function formalism,
implemented in branching and percolation processes on complex
networks. This approach allows to study not only the steady state but
also the temporal evolution of the process. First, we derive the
general equations for the case of unlimited vaccines and then we
explain the effect of the depletion of the vaccines.  Denoting the
fraction of susceptible, infected, vaccinated and recovered
individuals at time $t$ by $S(t)$, $I(t)$, $V(t)$ and $R(t)$,
respectively, the EBCM approach lies on describing the evolution of
the probability that a randomly chosen node is susceptible. In order
to compute $S(t)$, a link is randomly chosen and then a direction is
given, in which the node in the target of the arrow is called the {\it
  root node}, and the base is its neighbor, called {\it base node}.
We denote $\theta_t$ to the probability that at time $t$, the base
node does not transmit the disease to the root node and neither induces
the immunization of the root node. In this approach, the state of the
base node can not be affected by the root node, so that we can treat the
state of the root’s neighbors as independent
\cite{volzsir,miller_11,Val_12}. A node remains as susceptible if none
of its $k$ neighbors cause its infection or immunization, then the fraction of individuals in
the susceptible state at time $t$ is given by

\begin{equation}\label{SS}
S(t)=\sum_k P(k)\theta_t^k= G_0(\theta_t),
\end{equation}
where $G_0(x)=\sum_{k=k_{min}}^{k_{max}} P(k)x^k$ is the generating
function of the degree distribution, $P(k)$, of the network \cite{New_03}. To compute $\theta_t$ we have to take into account all the possible
states of the base node. Suppose an edge that connects the root node
and the base node. Then, this edge has not been used yet to infect or vaccinate
the root node if the base node is

\begin{itemize}

\item in state S, with probability  $\Phi_S$.
\item infected but did not spread the disease to the root node, nor
  induced the immunization of the root node, which is expressed by
  $\Phi_I$.
\item in state R but during the time it was infected, it did not
  propagate the disease to the root node, nor induced vaccination
  to the root node. This probability is denoted by $\Phi_R$.
\item  vaccinated or immunized, with probability $\Phi_V$.
\end {itemize}

We summarize these probabilities in Table~\ref{Tabbase}.

\begin{table}
\begin{center}
    \begin{tabular}{ | c | c |}
      \hline
      Quantity & Possible states of the base node \\ \hline
    $\Phi_S$ & Susceptible \\ \hline
    $\Phi_I$ & Infected and did not infect nor did it induce the vaccination of the root node  \\ \hline
   $\Phi_R$ & Recovered and did not infect nor did it induce the vaccination of the root node   \\ \hline
    $\Phi_V$ & Vaccinated \\
      \hline
    \end{tabular}
    \caption{Probabilities for the state of the base node, which is a
      neighbor of the root node, in the edge-based compartmental
      model.}\label{Tabbase}
\end{center}
\end{table}

\begin{figure}[ht]
  \begin{center}
    \includegraphics[width=0.7\textwidth]{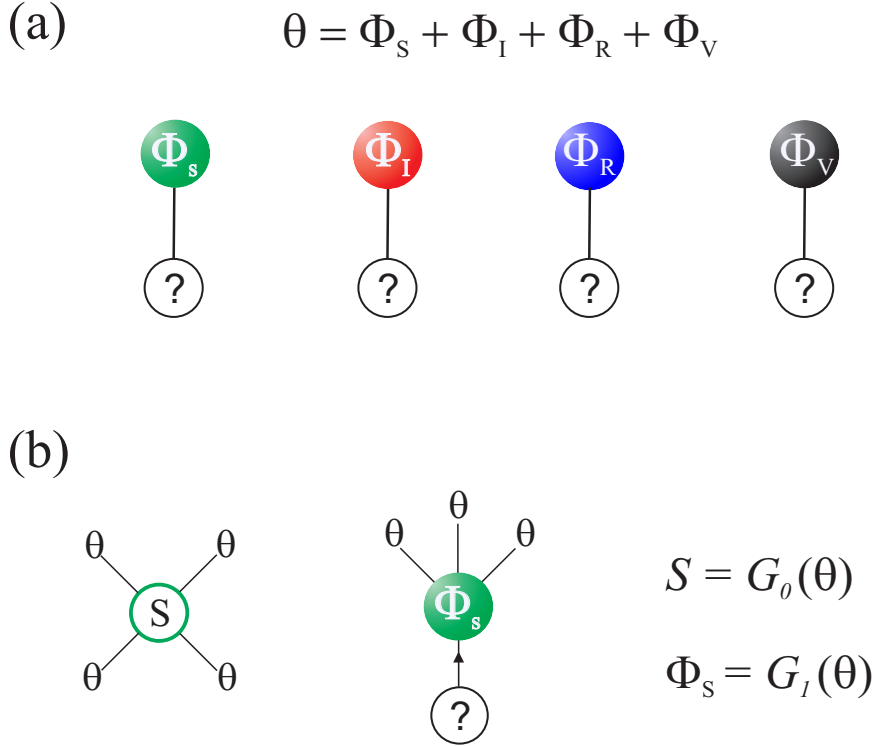}
  \end{center}
  \caption{Diagram showing the relations between the variables used in
    the compartmental model. The aim is to calculate the probability
    that the root node, denoted by a question mark, is
    susceptible. (a) The neighbor of the root node, called base node,
    does not spread the disease to the root, nor induce the vaccination
    of the root node with probability $\theta$. Shown are all the
    possible states of the base node. (b) A node is susceptible if the
    disease does not spread through its $k$ links, and if none of its
    $k$ partners induce its vaccination. We consider that the root
    node can not change the state of the base node, thus the latter
    node is susceptible if it does not get infected through its $k-1$
    links, and if none of its $k-1$ neighbors other than the root, cause its immunization.
  }
  \label{MillerScheme}
\end{figure}

In Fig.~\ref{MillerScheme} a) we demonstrate the configurations of the root and the base node.

Thus, accounting all these cases (see Fig.~\ref{MillerScheme} b),
$\theta_t$ is given by

\begin{equation}\label{Eqtheta}
  \theta_t=\Phi_S(t)+\Phi_I(t)+\Phi_R(t)+\Phi_V(t).
\end{equation}

Similar to $S(t)$ in Eq.~(\ref{SS}), we can write an expression for
$\Phi_S(t)$ (see Fig.~\ref{MillerScheme} b). The neighbor of the root
node, with degree $k$, is in state S if the disease does not spread
through its $k-1$ links and if none of its $k-1$ neighbors, omitting
the root node, do not induce its vaccination. Recall that the edge coming
from the root node is not considered. Hence the probability that the
base node is susceptible at time $t$ is $\theta^{k-1}_t$ and thus,

\begin{equation}\label{PhiS}
  \Phi_S(t)=G_1(\theta_t),
\end{equation}
where $G_1(x)=\sum_{k=k_{min}}^{k_{max}} k P(k)/ \langle k \rangle
x^{k-1}$ is the generating function of the excess degree distribution
of the network and $\langle k \rangle$ is the average degree
\cite{New_01}. The evolution equations that describe the process for
unlimited vaccination, i.e. $V_L=1$, are (see Appendix A for detailed
derivation),

\begin{flalign}\label{EqMiller}
  \Delta \theta_{t+1}&= -\big[\omega+(1-\omega)\beta\big] \Phi_I(t)\\ \Delta
  \Phi_S(t+1)&=G_1(\theta_{t+1})-G_1(\theta_{t})\nonumber\\ \Delta
  \Phi_I(t+1)&=-\big[\omega+(1-\omega)\beta\big] \Phi_I(t)-C_{\beta} \Delta
  \Phi_S(t)+(1-\Omega)C_{\beta}\Delta \Phi_S(t-t_r).\nonumber\\
  \Delta \Phi_V(t+1)&= - C_{\omega}\Delta\Phi_S(t)  .\nonumber
 \end{flalign}

In the first equation, $\theta_t$ decreases if the base node is
infected at time $t$, and induces the vaccination of the root node
with probability $\omega$, or if it spreads the disease to the root,
with probability $(1-\omega)\beta$. Note that $\Phi_I$ takes into
account that the base node and the root node had no prior
interaction. The second equation represents the evolution of the
probability that the base node is in state S, which is the finite
difference of $\Phi_S$ (see Eq.~(\ref{PhiS})). The third equation is a
bit more complicated. The root node has, with probability $\Phi_I(t)$,
an infected neighbor at time $t$ that did not induce its immunization
or caused its infection. This probability changes if the infected base
node causes the immunization of the root node or infects it, which is
reflected in the first term. In the second term we have the
susceptible individuals that become infected at time $t$ and will be
in state I in the next time step. Unlike \cite{Val_12}, where the
authors used the EBCM to solve the classical {\it SIR} model, this
term does not account all the variation of $\Phi_S(t)$, since a
fraction of the susceptible individuals go to state V. The fraction of
the nodes in state S that go to state I is weighted with the factor
$C_{\beta}=(1-\omega)\beta/\big[\omega+(1-\omega)\beta\big]$, which is
the probability that the disease spreads through a link. The last term
in the 3rd equation takes into account the susceptible individuals
that got infected at $t_r$ time units earlier, and did not change the
state of the root node during this time. The probability that they do
not spread the disease or induce the vaccination to the root node
during this period of time is $1-\Omega$, where

\begin{equation}\label{Omegon}
  \Omega=1-(1-\omega)^{t_r}(1-\beta)^{t_r},
\end{equation}
and $(1-\omega)^{t_r}(1-\beta)^{t_r}$ is the probability that during
the period $t_r$ an infected base node did not infect nor induced the vaccination to the root node. Finally, the last
equation takes into account the immunized neighbors of the root
node. The variation of $\Phi_V(t)$ increases with time and is proportional to the negative change of $\Phi_S$. In this case the
factor $C_\omega=\omega/\big[\omega+(1-\omega)\beta\big]$, is the
probability that the link between the root node and the base node is
used to immunize. We explain these additional probabilities in Table~\ref{Tab3}.

After computing $\theta_t$ using Eqs.(\ref{EqMiller}), we can compute the
evolution of the fraction of susceptible, infected and
vaccinated individuals at time $t$ by,

\begin{flalign}\label{EqMiller2}
  \Delta S(t+1)&= G_0(\theta_{t+1})-G_0(\theta_t),\\
  \Delta V(t+1)&= - C_\omega \Delta S(t), \nonumber\\
  \Delta I(t+1)&= C_\beta \big(-\Delta S(t) +\Delta S(t - t_r)\big),\nonumber
\end{flalign}

\begin{table}
\begin{center}
    \begin{tabular}{ | c | c |}
      \hline
      Probability & Definition \\ \hline
    $\theta$ & The base node did not infect nor did it induce the vaccination of the root node \\ \hline
    $C_{\beta}$ & A susceptible node adopts the state I if its state change   \\ \hline
   $C_{\omega}$ & A susceptible node adopts the state V if its state change  \\ \hline
    $\Omega$ & A link is used to infect or vaccinate during $t_r$ units \\
      \hline
    \end{tabular}
    \caption{Probabilities that take into account the different interactions
      between the root node and base node.}\label{Tab3}
\end{center}
\end{table}

Notice than using these magnitudes we can compute the fraction of
recovered individuals, $R(t)=1-S(t)-I(t)-V(t)$. The derivation of
these equations is similar to Eqs. (\ref{EqMiller}). In the first
equation, the change in the fraction of susceptible individuals is the
finite difference of $S$ (Eq.(\ref{SS})). Next, in the second equation
the variation of the vaccinated individuals is proportional to the
change in the susceptible individuals. This is since $\Delta S \leq
0$, and the factor $C_\omega$ takes into account the transition from
state S to V. In the third equation, the change in the fraction of
infected individuals is also proportional to the variation of the
susceptible individuals, but here the factor $C_\beta$ is related to
the transition from state S to I. Hence, $- C_\beta \Delta S(t)$ is
the fraction of new infected individuals at time $t$. On the other
hand, $- C_\beta \Delta S(t - t_r)$ is the fraction of individuals
that got infected $t_r$ temporal units earlier. Thus, this fraction
represents the individuals that move to state R at
time $t$, and hence contribute negatively to the fraction of infected
individuals.

The set of equations (\ref{EqMiller}) and (\ref{EqMiller2}) describes
the temporal evolution of the process with unlimited vaccines
($V_L=1$). Now we assume that we have a limited amount of vaccination
units, lower than the number of individuals in the system. Thus we
impose a limit $V_L$ as the maximal fraction of vaccinated
individuals.

The evolution of the system is the same as the unlimited case until
$V(t)$ reaches the vaccination limit, $V_L$. At this point there is no available vaccines and thus the vaccination probability becomes
zero, allowing the disease to spread without barriers. Hence the
equations should be iterated normally until $V(t)=V_L$, and then
setting $\omega=0$ for the rest of the process. Nevertheless, since in
Eq. (\ref{EqMiller2}), $V(t)$ changes by finite increments, it is
unlikely that $V(t)$ matches with $V_L$ exactly. Thus it is not clear
when iterating the equations, the precise moment at which the
immunization process has to be stopped. We explain
in Appendix B the procedure that has been performed to solve this problem and to reproduce exactly
the results from the computational simulations.

\section{Results}
\begin{figure}
  \begin{center}
    \includegraphics[width=0.48\textwidth]{TIME_SUSCEPTIBLES_ER_km10_w_045_vl_05_beta_0168_tr_3_rep.eps}  \hspace{0.2cm}
    \includegraphics[width=0.48\textwidth]{TIME_INFECTED_ER_km10_w_045_vl_05_beta_0168_tr_3_rep.eps}
    \includegraphics[width=0.48\textwidth]{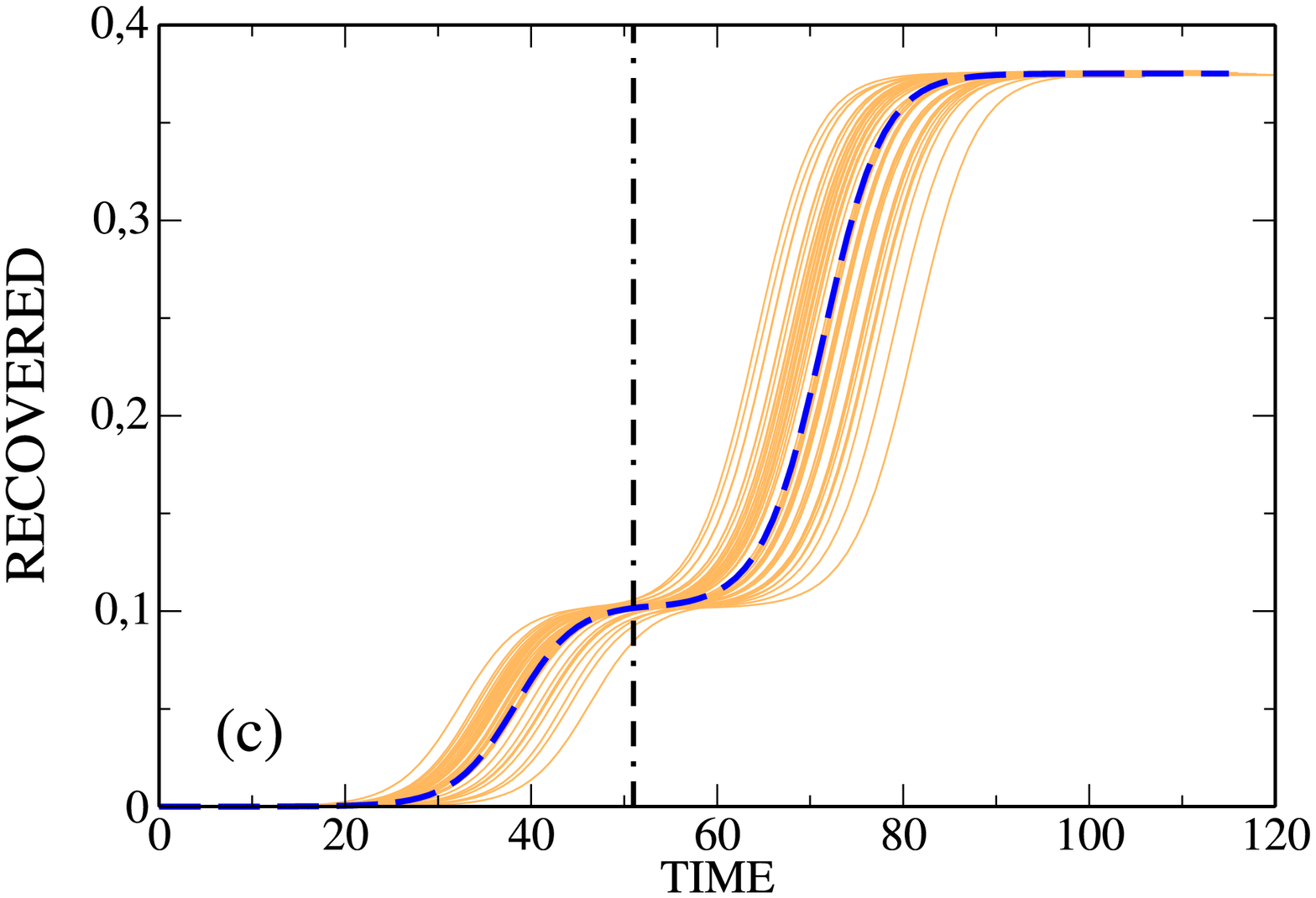} \hspace{0.2cm}
     \includegraphics[width=0.48\textwidth]{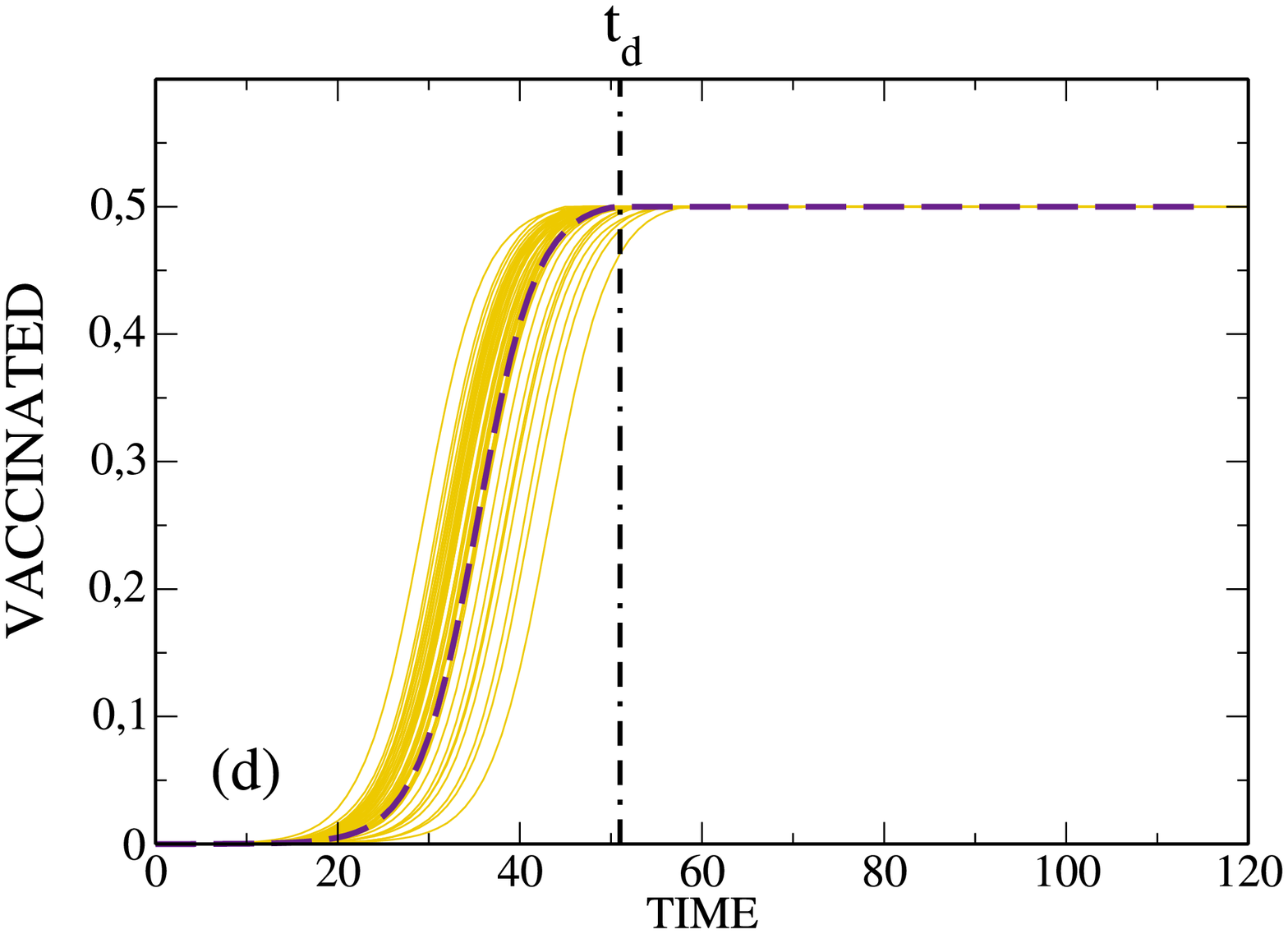}
      \end{center}
\caption{Temporal evolution of the SIR model with local vaccination
  for an $ER$ network with $\langle k \rangle = 10$, for
  $\omega=0.45$, $V_L=0.5$, $t_r=3$ and $\beta=0.168$. The solid lines
  represent different stochastic simulations, and the dashed lines
  are the theoretical results obtained from the EBCM,
  Eqs.~(\ref{EqMiller2}). The black dotted-dashed lines indicate the time at which the
  vaccines are depleted, denoted by $t_d$. We can see the very good
  agreement between the simulations and the theory.}\label{TEMPO1}
\end{figure}

We perform stochastic simulations of the localized and limited vaccination SIR model
over single networks with $N=10^6$ nodes, whose degree distribution is Erd\H{o}s R\'{e}nyi (ER) with an
average degree $\langle k \rangle=10$. The networks are built using the
Molloy-Reed algorithm \cite{Mol_01}.

\subsection{Temporal Evolution}

To demonstrate the validity of the theoretical formalism, in
Fig. \ref{TEMPO1} we show simulations and theory of the temporal evolution of the process for an
infection probability $\beta=0.168$, a vaccination probability
$\omega=0.45$, a vaccination limit $V_L=0.5$, and a recovery time
$t_r=3$. The dashed lines are the theoretical results from the EBCM
described in the previous section, while the solid lines represent
different realizations of the stochastic simulations. We can see the
excellent agreement between the theoretical equations
(\ref{EqMiller2}) and the simulation results.

It can be seen from Fig. \ref{TEMPO1} that initially, as the fraction
of infected and vaccinated individuals increase with time, the
fraction of susceptible individuals decreases. Each infected node
reaches the state R after $3$ units of time and consequently the
fraction of recovered individuals increases with time. In the
classical SIR model, the fraction of infected individuals reaches a
maximum and then decreases, but in our case of limited vaccines, after
a specific time the behavior of the curves changes. At this time the
system runs out of vaccine units and thus no more
individuals can be immunized against the disease. At this point the
fraction of susceptible nodes shows a plateau, since as seen in
Fig.~\ref{TEMPO1}(b), the epidemic almost vanished. and there are
only few infected individuals that can infect the susceptible people.
This plateau, which is also observed for the recovered individuals
(Fig.~\ref{TEMPO1}(c)), seems to indicate that the system begins to
stabilize, since the magnitudes change slowly with time. However, when
suddenly the vaccines are exhausted, the fraction of infected
individuals starts to increase again, reaching a second peak
(Fig.~\ref{TEMPO1}(b)) that could be even higher than the first
one. This increase obviously cause a further decrease and increase of
the susceptible and recovered individuals respectively. Finally the
disease starts to fade away as the fraction of infected individuals
decreases, then the system reaches the steady state and all
the magnitudes stabilize.

From now on, in the rest of the manuscript, our results will be based only
from the theoretical equations, since we find excellent agreement
(Fig. \ref{TEMPO1}) with the stochastic simulations.

Next we will show the temporal behavior of the process for different
values of the infection probability, $\beta$. In the standard SIR
model, without vaccination, there is a critical value $\beta_c$ below
which there is no epidemic. This value satisfies the equality
$T_c=1/\kappa$ \cite{Coh_01}, where
$T=1-(1-\beta)^{t_r}$ is the transmissibility, the effective
probability of contagion, and $\kappa=\langle k^2\rangle/\langle k
\rangle$ is the branching factor of the degree distribution of the
network. $\langle k\rangle$ and $\langle k^2\rangle$ are the first and
second moments of the degree distribution respectively. From this
relation, in which $T_c\equiv T(\beta_c)$, the critical infection
probability can be obtained as
$\beta_c=1-(\kappa-2)/(\kappa-1)$ \cite{New_05}. In the present model
of limited dynamical vaccination, the relation between $T_c$ and
$\kappa$ holds but in this case the transmissibility depends also on
the vaccination probability $\omega$ (see Appendix D for the
expression of $T$). Closed expressions of $\beta_c$ for $t_r>1$ are
quite complicated, however for $t_r=1$ the critical infection
probability is simply
$\beta_c=1-(\kappa-2)/\big((1-\omega)(\kappa-1)\big)$. We see that
this probability depends on $\omega$ but not on the vaccination limit
$V_L$. Besides $\beta_c$, in our model there are also specific values
of $\beta$ associated with dramatic changes in the behavior of the
magnitudes at the steady state. Unlike $\beta_c$, these values depend
on the number of immunization units. Next we will show how the
magnitudes evolve with time when $\beta$ is one of these specific
values.

\begin{figure}
  \begin{center}
    \includegraphics[width=0.48\textwidth]{Temporal_ER_km_10_w_045_nu_04_beta_025809.eps} \hspace{0.2cm}
    \includegraphics[width=0.48\textwidth]{Temporal_ER_km_10_w_045_nu_04_beta_02581.eps} 
    \includegraphics[width=0.48\textwidth]{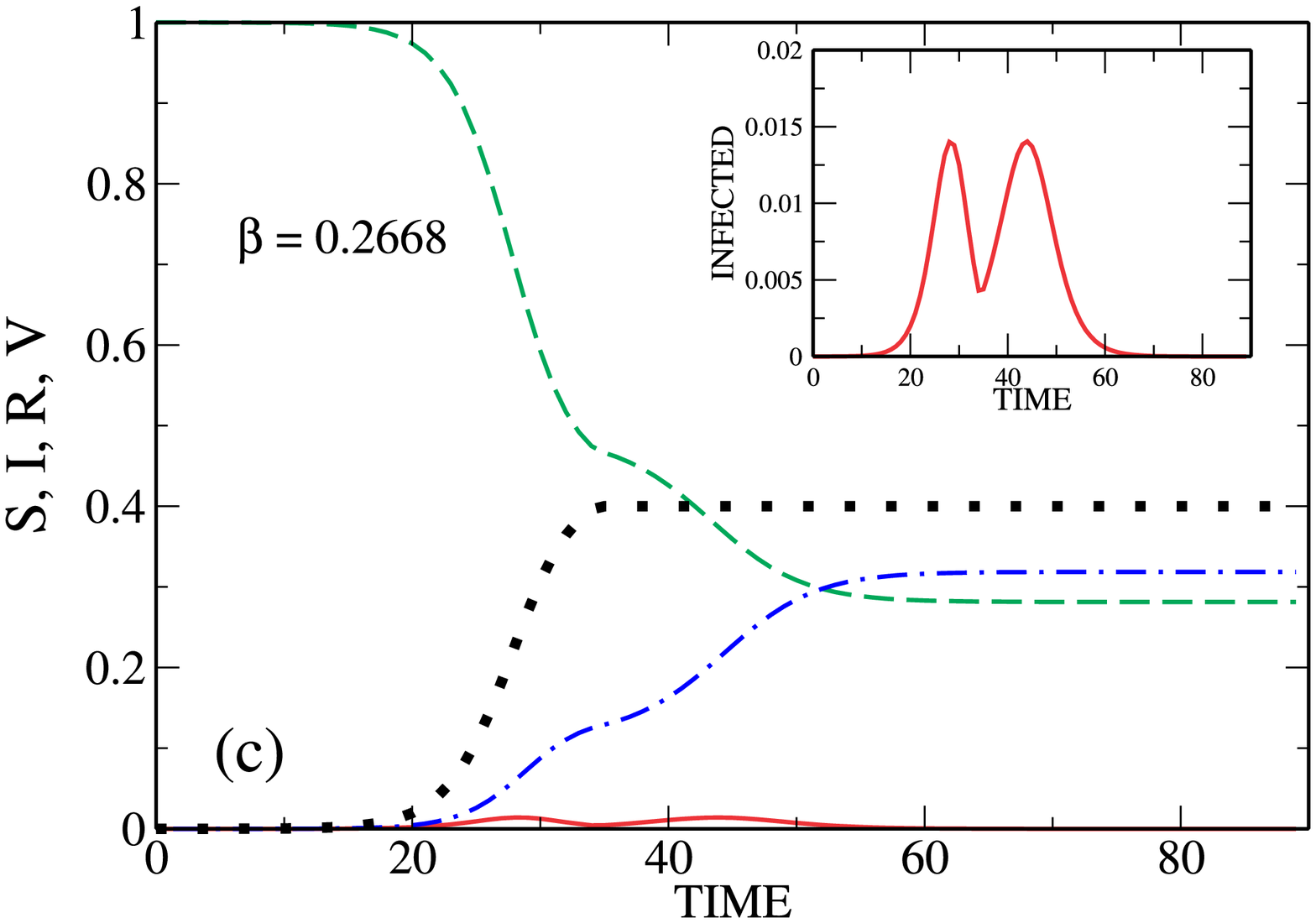} \hspace{0.2cm}
    \includegraphics[width=0.48\textwidth]{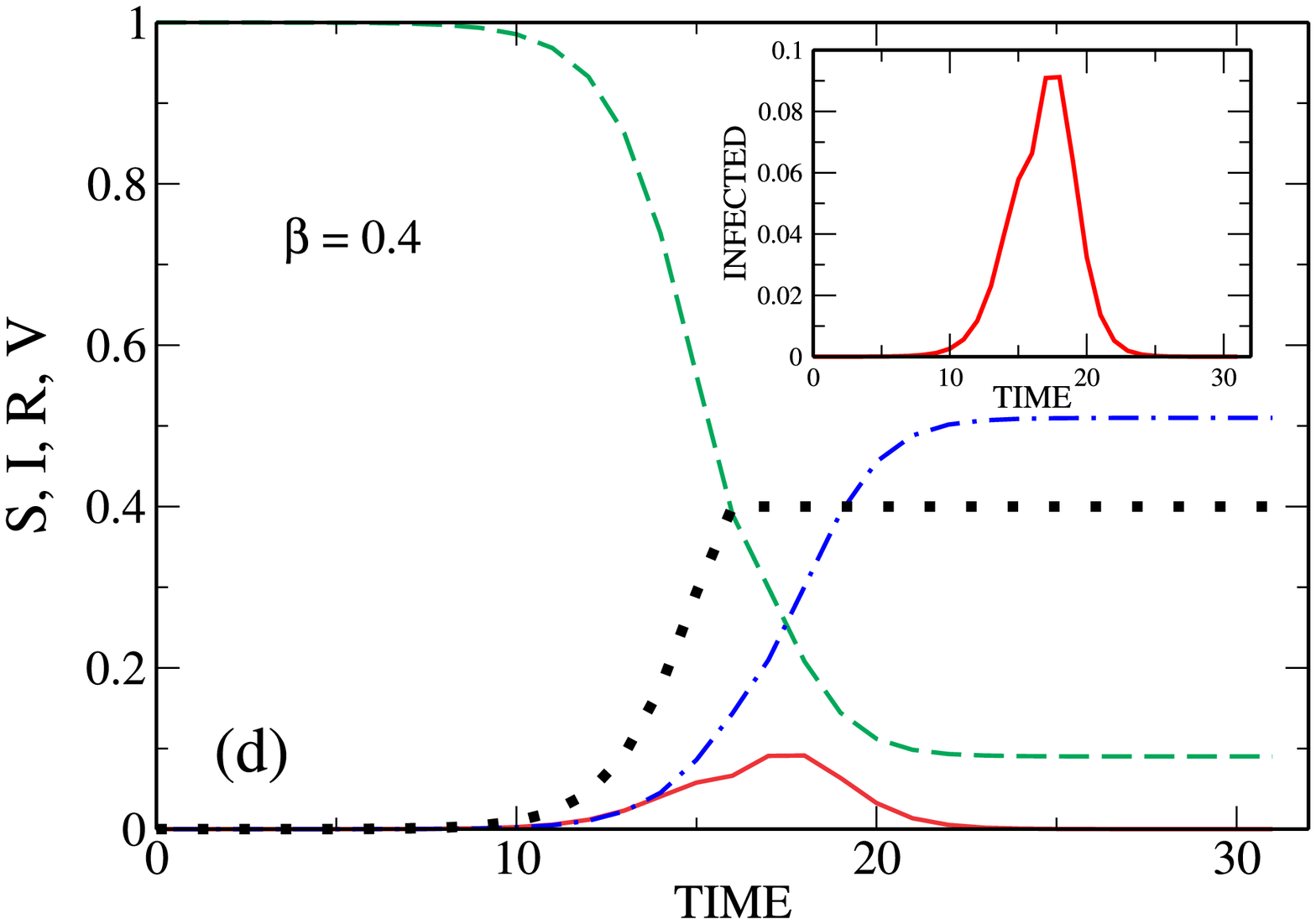}
      \end{center}
\caption{Temporal evolution of the model for a $ER$ network with $\langle k \rangle
  = 10$, for $\omega=0.45$, $V_L=0.4$ and $t_r=1$. The curves
  represent susceptible nodes (\protect\verdash), infected
  (\protect\redsolid), recovered (\protect\bluedashdotted), and
  vaccinated (\protect\bckdotted). (a) $\beta=0.25809$, (b)
  $\beta=0.2581$, (c) $\beta=0.2668$ and (d) $\beta=0.4$. The insets show a magnified image of the temporal fraction of infected
  individuals.}\label{TEMPO2}
\end{figure}

Fig. \ref{TEMPO2} exhibits the temporal evolution of all magnitudes
for $\omega=0.45$, $V_L=0.4$, $t_r=1$ and for several values of
$\beta$. For this particular set of parameters $\beta_c=0.1818$,
nevertheless in this figure we will focus on another important value greater than $\beta_c$,
which we call $\beta^\ast$. In Fig. \ref{TEMPO2} (a) $\beta=\beta^\ast=0.25809$, and

we observe that in this case the fraction of infected individuals shows only one peak, and also the other magnitudes show a standard behavior. However if the
infection probability is just a little higher, $\beta=0.2581$,
suddenly two peaks appear in the fraction of infected nodes, as seen
in the inset of Fig. \ref{TEMPO2} (b). Similar to Fig. \ref{TEMPO1},
when the fraction of vaccinated individuals reaches the vaccination
limit, in this case $V_L=0.4$, the fraction of susceptible and
recovered nodes enter in a quasi-stationary state, in which they
barely change. At this stage there is a negligible number of infected
individuals, however since the vaccination stops, there are enough infected-susceptible pairs to start a second outbreak. This causes the second
peak and a further decrease in the number of susceptible individuals, as well as a further increase in the fraction of recovered
nodes. Thus, at $\beta^\ast$ the steady state
experiences an abrupt transition with $\beta$. We will see later in
Figs. \ref{SRV_beta} and \ref{Diagram} how the nature of the abrupt
transitions depends highly on the limited vaccines units, $V_L$.

In Figs. \ref{TEMPO2} (c) and (d), for larger values of $\beta$ we observe that
the two peaks start to get closer, until eventually they start to fuse
together forming a single peak as in Fig. \ref{TEMPO2} (a), but much
higher. On the other hand, for larger $\beta$ values the curves become smoother since there is only a single outbreak.

\subsection{Steady State}

\begin{figure}[h]
  \begin{center}
     \includegraphics[width=0.47\textwidth]{V_w_045_ER_km_10_new.eps}
      \hspace{0.5cm}
      \includegraphics[width=0.47\textwidth]{R_w_045_ER_km_10_new.eps}
     \includegraphics[width=0.47\textwidth]{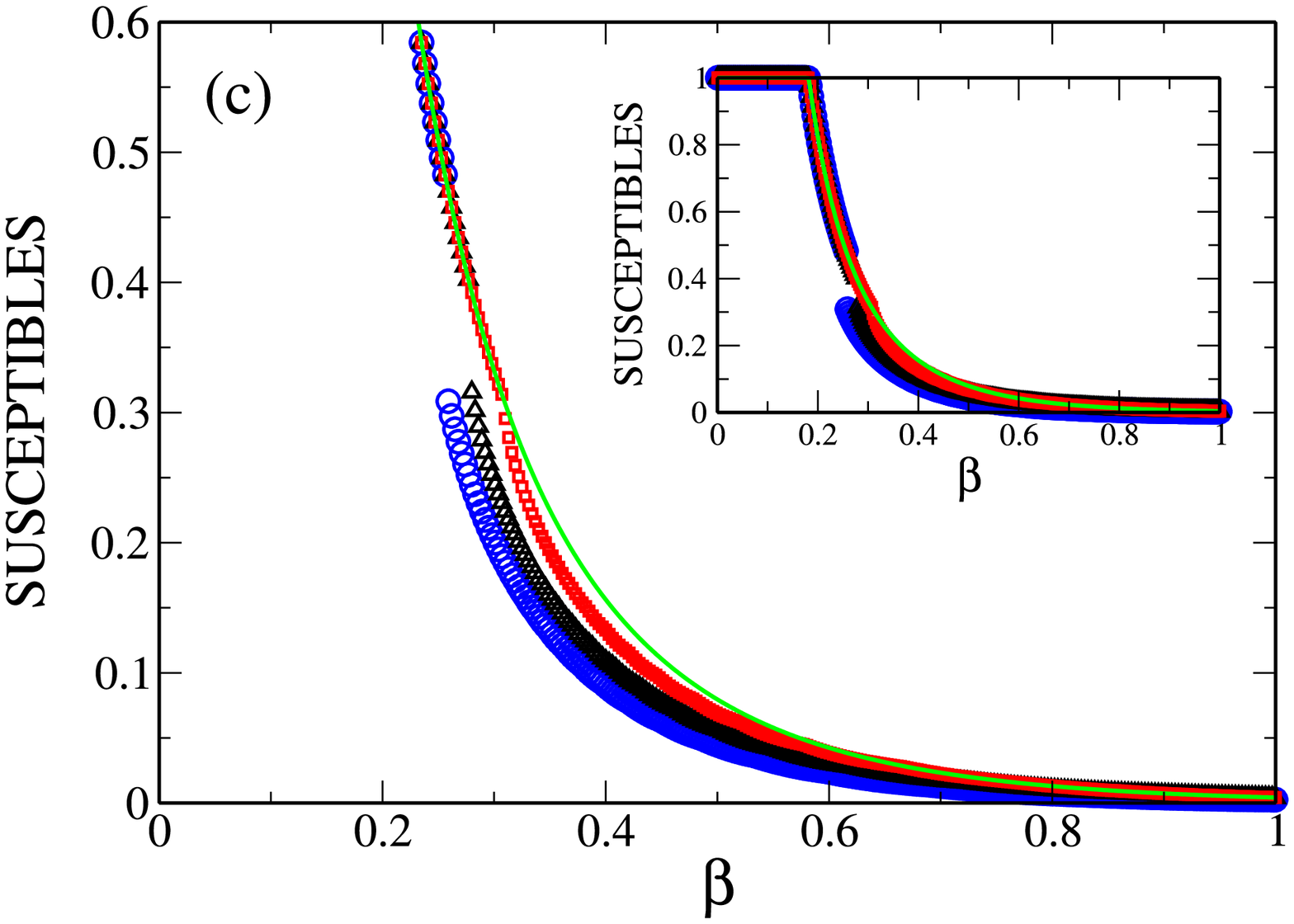}
     \hspace{0.5cm}
     \includegraphics[width=0.47\textwidth]{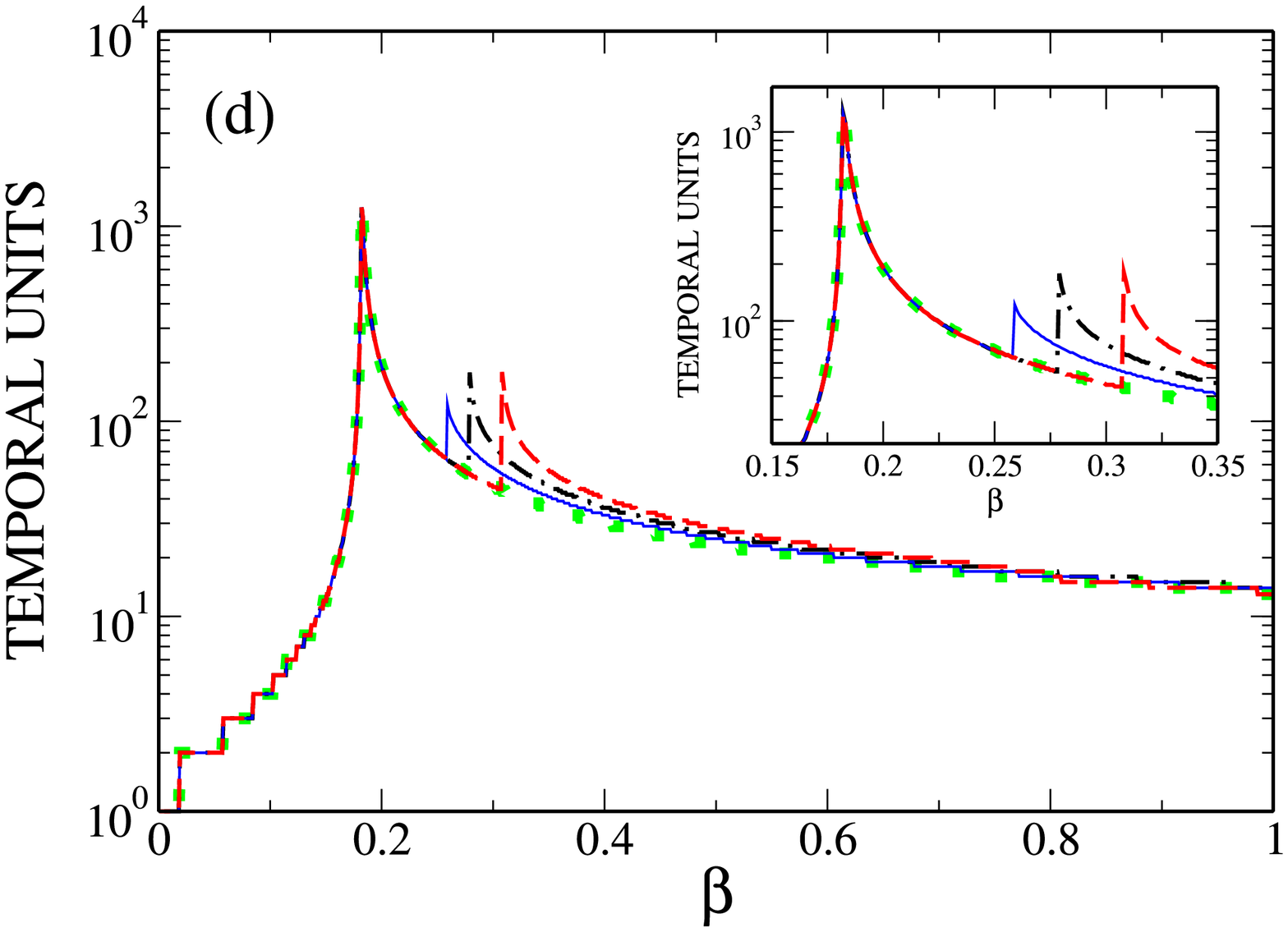}
  \end{center}
  \caption{Fraction of (a) vaccinated, (b) recovered and (c) susceptible
    individuals at the steady state as a function of the infection
    probability $\beta$. The degree distribution is ER with
    $\langle k \rangle=10$, $k_{min}=0$, $k_{max}=40$ and the
    recovery time is $t_r=1$. The vaccination probability is
    $\omega=0.45$ and the vaccination limits are: $V_L=1$
    (\protect\grsolid), $V_L=0.5$ (\protect\rojsqII), $V_L=0.45$
    (\protect\blcktrgII), $V_L=0.4$ (\protect\azcircII). The vertical
    dashed lines indicate the values of $\beta^\ast$ and $\beta^\dag$
    for $V_L=0.4$ and $V_L=0.5$. In (c), we show in the inset the
    full curve of susceptible nodes. In (d) we compute
    theoretically the time it takes the process to reach the steady
    state. Since the peaks are very close we show in the inset the curves near criticality for a
    better visualization. In this figure $V_L=1$
    (\protect\greendotted), $0.5$ (\protect\reddashed), $0.45$
    (\protect\blackdashdotted), $0.4$
    (\protect\bluesolid)}\label{SRV_beta}
\end{figure}

To understand how the results are affected by the infection
probability $\beta$, in Fig.~\ref{SRV_beta} we show the fraction of
vaccinated, recovered and susceptible individuals in the steady state
as a function of $\beta$ for a fixed vaccination probability
$\omega=0.45$, $t_r=1$ and for different values of limited vaccines
$V_L$, for an ER network with $\langle k \rangle=10$. The green solid
lines represent the case of unlimited vaccines, .i.e., $V_L=1$.  In
Fig.~\ref{SRV_beta} (a) we see that for increasing $\beta$, the
fraction of vaccinated individuals increases, since more infected
individuals means more susceptible neighbors to
immunize. Nevertheless, we observe that this curve reaches a maximum
and then decreases for larger values of $\beta$. This can be
understood as follows. When the probability of infection becomes high, the majority of neighbors of an infected node get infected instead of being immunized,
and thus there is a decrease in the fraction of vaccinated individuals
\cite{Alvarez_2018}. It is important to point out that the existence
of this maximum is highly influenced by the vaccination probability $\omega$ and
the topology of the network. In Appendix E (Fig.~\ref{Vac_SF}) we show
the fraction of vaccinated and recovered individuals at the steady
state for networks with a heterogeneous power law degree distribution, and for different values of
$\omega$. 

Next we observe what happens if we impose a limit on the fraction of
available vaccines. For $V_L=0.5$, represented by the red squares, we
see in Fig.~\ref{SRV_beta} (a) a plateau between two values of
$\beta$. This happens since V can not surpass the vaccination
limit. {The lower} of these values is $\beta^\ast$, which we introduce in
Fig.~\ref{TEMPO1}, and we call the other $\beta^\dag$, which is
greater than $\beta^\ast$ (see Fig.~\ref{SRV_beta} (a)). Between these two
values is the range of $\beta$ for which V reaches its limit value,
$V_L$. Now we ask what is the effect of the vaccination limit {on} the
fraction of recovered individuals. In Fig.~\ref{SRV_beta} (b) we observe
that between $\beta^\ast$ and $\beta^\dag$, denoted by the vertical
dashed lines, the curves of recovered fraction show significantly
increased values compared to the case of unlimited vaccines. In this region at
some value, $\beta^\ast$, the vaccine units are exhausted, and then the
disease spreads without barriers, affecting a great number of
individuals that could not be immunized.

For example, for $V_L=0.45$, represented in Fig.~\ref{SRV_beta} by black triangles, we observe a similar
behavior. In in this case $\beta^\dag=1$, but an interesting phenomena
takes place at $\beta^\ast$ where a discontinuous jump occurs due to
the shortage of vaccination units. This abrupt transition, in Fig.~\ref{SRV_beta} (b) and (c), can be
understood from Figs.~\ref{TEMPO2} (a) and (b). Below
$\beta^\ast$ there is a single outbreak, while above it a second
outbreak causes the abrupt jumps observed in
Fig.~\ref{SRV_beta}. It is expected therefore that for $V_L=0.4$, an
even smaller supply of immunization units, the condition of the
population becomes worse, as seen in Fig.~\ref{SRV_beta}. The curve
with blue circles shows that there is no $\beta^\dag$, and thus for any
$\beta>\beta^\ast$ the fraction of infected is significantly
higher, compared to the case of unlimited vaccines. The dashed lines that denote these points indicate the
emergence of a second peak of infection, as can be seen in
Figs.~\ref{TEMPO2} (b) and (c).

In Fig.~\ref{SRV_beta} (c) we show the fraction of
susceptible individuals, which decrease with $\beta$ and also show a
discontinuous jump at $\beta^\ast$, associated with $V_L$.

 In Fig.~\ref{SRV_beta} (d) we compute the time it
takes the process to reach the steady state, as seen
  in Fig.~\ref{TEMPO2}. Processes near the transition point, usually
have longer duration times compared to
those far from this point, as found for example in
the process of cascading failures \cite{Par_02,Zhou14}. Taking this
into consideration, we can see that for $V_L=1$,
represented in this figure by a green dotted line, there is only one
peak at $\beta_c=1/(\langle k \rangle (1-\omega))$
\cite{Alvarez_2018}, which is associated with a continuous phase
transition. This peak is seen for all $V_L$, since
the critical point does not depend on $V_L$. However, for $V_L=0.4$,
$V_L=0.45$ and $V_L=0.5$, represented respectively by a blue solid
line, a dash-dotted black line, and a red dashed line, we observe
another peak at longer times, which is located at $\beta^\ast$,
which depends on $V_L$. Consequently, if we return to
Fig.~\ref{TEMPO2}, where we show the temporal evolution for $V_L=0.4$,
we see indeed that for $\beta^*=0.2581$ (Fig.~\ref{TEMPO2} (b)) the
process takes much {longer} time to reach the steady
state compared to $\beta=0.25809$ (Fig.~\ref{TEMPO2}
(a)).

Thus, we can infer that similar to $\beta_c$, the probability of
infection $\beta^\ast$ denotes a transition point, that separates a
region in which the immunization strategy stops the spreading of the
disease, and another in which the vaccination units are insufficient
to stop it. On the other hand, we do not observe a peak at
$\beta^\dag$, indicating that this is not a transition point. Instead,
this point denotes a crossover between the regime of insufficient
vaccines and a regime in which the immunization units can not be used
completely, since the probability of infection is too high. Moreover
we recall that the existence of $\beta^\dag$ is related to the
topology of the network, see Appendix E (Fig.~\ref{Vac_SF}). To see
also how the curves of Fig.~\ref{SRV_beta} behave for a different recovery
time $t_r$, see Fig.~\ref{SRV_beta_tr3} in appendix E where we show the steady state for $t_r=3$ and for $V_L=0.4$, the vaccination limit used in
Fig.~\ref{TEMPO1}.

\begin{figure}
  \begin{center}
      \includegraphics[width=0.47\textwidth]{Recovered_vs_w_beta_02_new.eps}    
      \hspace{0.5cm}   
      \includegraphics[width=0.47\textwidth]{Recovered_vs_w_beta_05_new.eps}
      \includegraphics[width=0.47\textwidth]{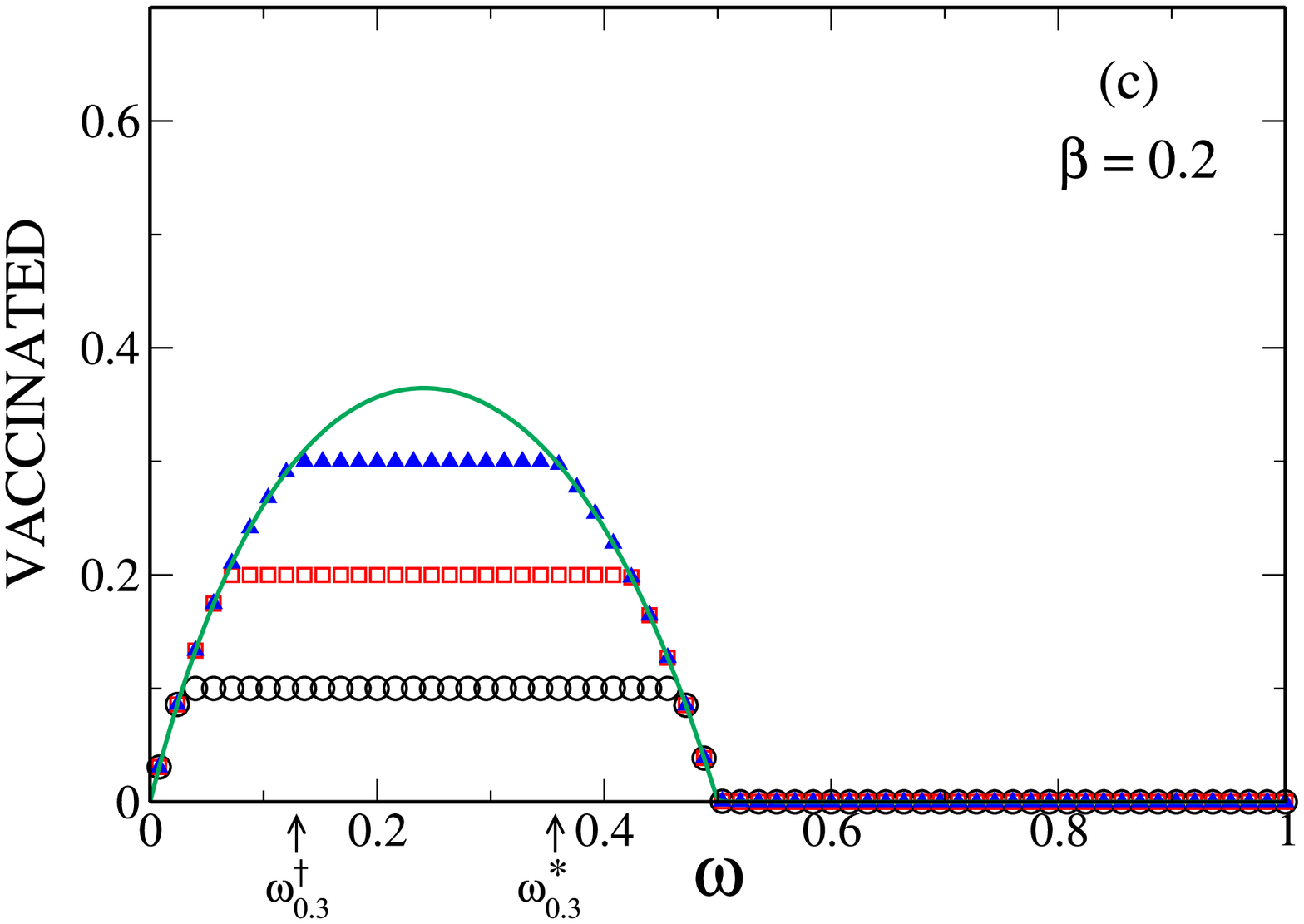}
       \hspace{0.5cm}
     \includegraphics[width=0.47\textwidth]{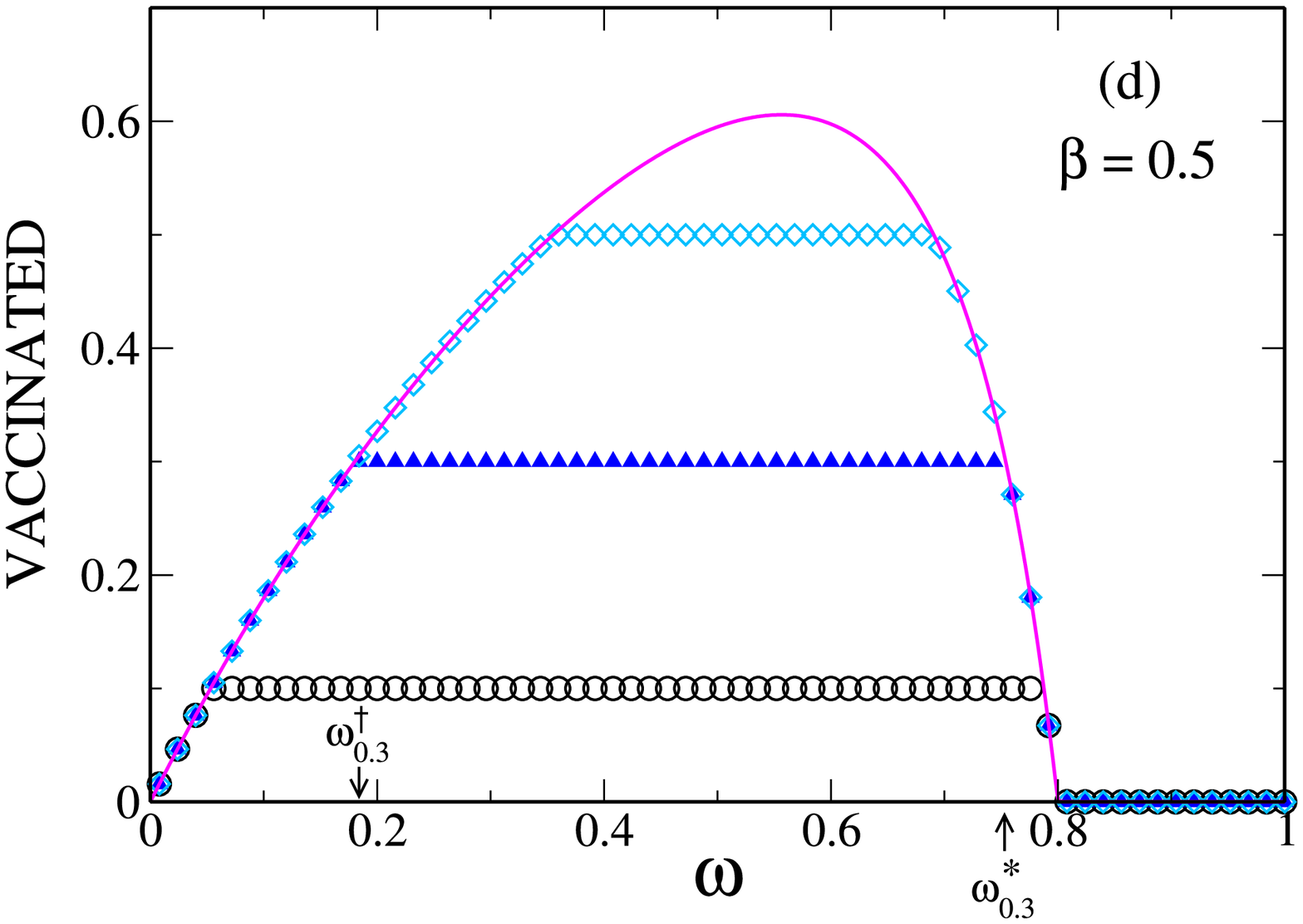}
  \end{center}
  \caption{Fraction of vaccinated and recovered individuals as a
    function of $\omega$ for fixed $\beta$ values and for $t_r=1$. The
    curves represent different vaccination limits: $V_L=0.1$
    (\protect\blackcircle), $V_L=0.2$ (\protect\redsquare), $V_L=0.3$
    (\protect\bluetriangle), $V_L=0.4$ (\protect\verdecitosolid),
    $V_L=0.5$ (\protect\skydiamond) and $V_L=0.7$
    (\protect\magsolid). In (a) and (c) the infection probability is
    $\beta=0.2$, while in (b) and (d) $\beta=0.5$. Note that for
    $\beta=0.5$ ((b) and (d)), the number of recovered and vaccinated
    individuals are larger compared to the case $\beta=0.2$, ((a) and
    (c)). The dashed vertical lines indicate the jumps and the arrows the values of $\omega^\ast$ and
    $\omega^\dag$ for $V_L=0.3$. These values are different for
    $\beta=0.2$ and $\beta=0.5$.}\label{RV_w}
  \end{figure}

Next we fix the infection probability $\beta$ and analyze how the
magnitudes at the steady state change with the
vaccination probability $\omega$. In Fig.~\ref{RV_w} we show the
fraction of recovered and vaccinated individuals as a function of
$\omega$ for different vaccination limits. First we focus
on the cases $V_L=0.4$ in Figs.~\ref{RV_w} (a) and
(c), and $V_L=0.7$ in Figs.~\ref{RV_w} (b) and (d). Similar to
Fig.~\ref{SRV_beta} the fraction of vaccinated individuals increases
with $\omega$ and reaches a maximum, after which it starts decreasing. This occurs because many of the
paths that would be used by the disease to spread, are blocked by
immunized individuals. Thus, since there are few people infected
there are fewer contacts around them to vaccinate. Furthermore, there
is a critical vaccination probability $\omega=\omega_c$ for which the
disease stops to propagate, since all the paths are completely blocked
due to vaccination. For $t_r=1$ we can show that $\omega_c=1-1/(\beta \langle k
\rangle)$ \cite{Alvarez_2018}. We see that for these values of $V_L$
the amount of vaccines is sufficient. Next we examine smaller values
of $V_L$, for which the system runs out of vaccines at some point.

 In all figures we observe that as $\omega$ increases from zero, the
 number of vaccinated individuals rises and the number of recovered
 individuals decreases, until a specific vaccination probability
 $\omega=\omega^\dag$, for which the vaccines are depleted. We see in
 Fig.~\ref{RV_w} (a) and Fig.~\ref{RV_w} (b) that after this point the
 fraction of recovered individuals has a lower decline rate, being
 practically insignificant for small values of $V_L$. For instance,
 from Fig.~\ref{RV_w} (b) it is clear that vaccination with probability
 $\omega=0.1$ or $\omega=0.7$ yield the same results. However,
 suddenly at $\omega=\omega^\ast$, a small increase in the fraction of immunized
 individuals can block many spreading paths of the disease, which
 results in a dramatic drop in the number of recovered
 individuals. This discontinuous jump is analogous to the behavior
 observed in Fig.~\ref{SRV_beta} at $\beta^\ast$, and one can easily
 relate $\omega^\dag$ to $\beta^\dag$.

 Thus, for a disease with $\beta$ fixed and for a fixed number of
 available vaccines $V_L$, based on the vaccination rate we can
 predict the number of infected individuals in the system when the
 epidemic comes to an end. Furthermore and very
   importantly, for a given $\beta$ and $V_L$, we can chose the optimal rate
   of vaccination, $\omega$, such that the fraction of infected be minimal or
   even zero.
 
The numerical values of $\beta^\ast$, $\beta^\dag$, $\omega^\ast$ and
$\omega^\dag$ can be calculated theoretically using the generating
functions formalism and branching theory \cite{Dun_01,New_03} (see
Appendix D for the derivation of the formula).

\begin{figure}
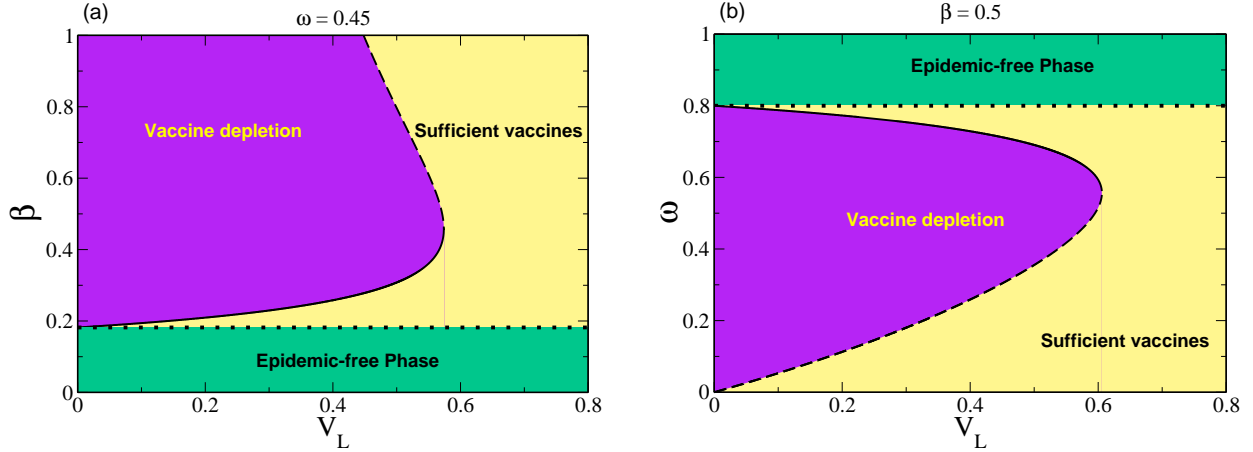

  \begin{center}
    \includegraphics[width=0.48\textwidth]{PDbeta.eps}
      \hspace{0.5cm}  
      \includegraphics[width=0.47\textwidth]{PDw.eps}
      
  \end{center}
  \caption{Phase diagram in (a) the plane $\beta$-$V_L$ with $\omega=0.45$ and in (b) the plane $\omega$-$V_L$ with $\beta=0.5$, for an
    ER network with $\langle k \rangle =10$ with $t_r=1$. The solid
    and dashed lines represent respectively $\beta^\ast$ and $\beta^\dag$ in (a), and $\omega^\ast$ and $\omega^\dag$ in
    (b). The dotted lines denotes the critical probabilities of
    infection, $\beta_c$ in (a), and $\omega_c$ in (b).}\label{Diagram}
\end{figure}

Finally in Fig.~\ref{Diagram} we show the model phase diagrams, which exhibit
different regions depending on the parameters. In Fig.~\ref{Diagram}
(a), where $\omega=0.45$, the solid curve in the $V_L-\beta$ plane
represents the values of $\beta^\ast$ for each $V_L$, while the dashed
curve represents $\beta^\dag$. This curves enclose the
vaccines-depletion region (in purple), where we see that for a
small vaccination limit and a high infection probability the system
runs out of immunization units. On the other hand, the region of
sufficient vaccines (in yellow), is characterized by (i) low
infection probabilities, for which little vaccines are needed to stop the
disease, and (ii) if $\beta$ and $V_L$ are both high, the vaccines do
not get exhausted. The reason for (ii) is that the disease
rapidly spreads before all the vaccines can be used. Finally the
dotted line represents the value of $\beta_c$, below which there is no
epidemic.

In Fig.~\ref{Diagram} (b) we fix $\beta=0.5$ and show the phase diagram in the plane
$V_L-\omega$. Here the solid and dashed curves represent $w^\ast$ and
$w^\dag$ respectively. Here we see that when $\omega$ increases the
depletion region becomes broader because more vaccines are
applied. However when $\omega$ further increases, the immunization
strategy gets more effective against the disease, and then a
smaller amount of vaccines is required to control the
epidemic. Furthermore when $\omega\geq \omega_c$, the disease can not
propagate since all the paths are blocked by immunized individuals.

 Using these phase diagrams we can learn how the regions of
 insufficient vaccines change with the available
 immunization resources in medical institutes, the infection
 probability, which depends on the disease, and the vaccination
 probability, which may depend on the medical workers.

 \section{Discussion}
 
In this manuscript we have explored the implications of a limited
number of vaccines in the $SIR$ model with local vaccination. We find
that at the steady state, there is a region of values of the infection
probability $\beta$, in which the medical institutions run out of
immunization units. This region is delimited by $\beta^*$ and
$\beta^\dag$ or, by $\beta^*$ and $\beta=1$ depending on the
vaccination limit $V_L$. We also find that $\beta^*$ is a transition
point, at which the curve of recovered individuals has a discontinuous
jump, whose height depends on $V_L$. This type of behavior, in
which a discontinuous transition is observed, has been seen when the
dynamics of propagation of epidemics is coupled with social processes
\cite{velasquez17,gomez16}. Furthermore, we analyze the temporal
evolution of the process close to $\beta^*$. We find that for $\beta
\gtrsim \beta^*$, the temporal evolution of the fraction of infected
individuals presents two peaks. When the disease is about to vanish
the vaccines are exhausted, and then the infection probability
$\beta^*$ is sufficiently large for an extremely small fraction of
infected individuals to cause a sudden second outbreak. On the
contrary, we observed that $\beta^\dag$ is not a transition point but
a crossover, and that its existence depends on the topology of the
network.

On the other hand we analyze the steady state of the process as a
function of $\omega$, finding other points of interest. One of them is $\omega^\dag$, below which the vaccination probability is too low to
use all the vaccines, thus the immunization units are not exhausted but
the epidemic is not effectively halted. Another point is
$\omega^\ast$, above which the vaccination probability is high enough
to control the epidemic with the available immunization resources, and
shows a discontinuous transition in the fraction of recovered
individuals. These results are of significant importance since the
vaccination probability is one of the few parameters that can be
controlled by the health institutes. Thus, $\omega$ can be chosen to
minimize the number of infected individuals or even halt the epidemic
in the primary stages, according to the available resources.

We solved the model using an EBCM, finding an excellent agreement with
the stochastic simulations. Also, we used the branching theory to find
the values of $\beta^*$, $\beta^\dag$, $\omega^*$ and
$\omega^\dag$. In future studies we will analyze different features of
the local vaccination model, as the immunization of first and second
neighbors of infected individuals. Thus, intending imitate more
accurately the ring vaccination strategy used against the Ebola
outbreak in Guinea during $2015$.
 

\section*{Acknowledgements}

SH thanks the Israel Science Foundation, ONR, the Israel Ministry of
Science and Technology (MOST) with the Italy Ministry of Foreign
Affairs, BSF-NSF, MOST with the Japan Science and Technology Agency,
the BIU Center for Research in Applied Cryptography and Cyber
Security, and DTRA (Grant no. HDTRA-1-10-1- 0014) for financial
support. LAB wish to thank DTRA (Grant no. HDTRA-1-10-1- 0014) for
financial support. MAD, LGAZ and LAB wish to thank to UNMdP and CONICET
(PIP 00443/2014) for financial support.

\appendix

\section{Edge-based compartmental model (EBCM)}

In this appendix we derive the set of equations (\ref{EqMiller}) using the EBCM. For simplicity
we will assume continuous time and rates $r_\beta$ and $r_\omega$ of
infection and vaccination respectively. Also the recovery time $t_r$
is replaced by a recovery rate $\gamma$. Once we derive the
equations of temporal evolution we can adapt them for
discrete time steps.

As we saw earlier the probability that the root node of the
network, selected at random, do not get infected or vaccinated through
a link by the base node is $\theta$, and satisfies
Eq. (\ref{Eqtheta}). This variable can change only if the link between the
root and the base node is used to infect or vaccinate. Since these
events occur with rates $r_\beta$ and $r_w$ respectively, thus

\begin{equation}\label{Dtheta}
\dot{\theta}=-(r_\beta+r_\omega)\Phi_I.
\end{equation}

Therefore, since $\Phi_S=G_1(\theta)$ then

\begin{equation}\label{DPhiS}
\dot{\Phi}_S=-(r_\beta+r_\omega)G^{'}_1(\theta) \Phi_I.
\end{equation}

On the other hand, a node is in state $V$ if is not susceptible and if
was more likely to receive a vaccine rather than be
infected, hence

\begin{flalign}\label{PhiV}
V&=\frac{r_\omega}{r_\omega+r_\beta}\big(1-G_0(\theta)\big),\\
\Phi_V&=\frac{r_\omega}{r_\omega+r_\beta}\big(1-G_1(\theta)\big).\nonumber
\end{flalign}

Similar to Eq.(\ref{DPhiS}) we can write

\begin{equation}\label{DPhiV}
\dot{\Phi}_V= r_\omega G^{'}_1(\theta) \Phi_I.
\end{equation}

Next we study the variation of $\Phi_R$, the probability that the base node is recovered and also, that during the time
it was infected did not cause the infection or the vaccination of the
root node. Since individuals recover with rate $\gamma$ hence

\begin{equation}\label{DPhiR}
  \dot{\Phi}_R=\gamma\Phi_I.
 \end{equation}

To obtain $\Phi_R$ first we have to rewrite Eq.(\ref{Dtheta}). The
probability that the disease or the vaccination spread through at
least one link to the root node is $1-\theta$ and thus

\begin{equation}\label{1-theta}
  \frac{d(1-\theta)}{dt}=(r_\beta+r_\omega)\Phi_I.
\end{equation}

Now combining Eqs. (\ref{DPhiR}) and (\ref{1-theta})

\begin{equation}
\frac{\gamma}{r_\beta+r_\omega}\frac{d(1-\theta)}{dt}=\frac{d\Phi_R}{dt}.
\end{equation}

Now integrating this equation and considering that $1-\theta$ and
$\Phi_R$ are negligible at the beginning of the process, then simply

\begin{equation}\label{PhiR}
\frac{\gamma}{r_\beta+r_\omega}(1-\theta)=\Phi_R.  
  \end{equation}

Combining Eqs. (\ref{Eqtheta}), (\ref{PhiV}) and (\ref{PhiR}) then

\begin{equation}
  \Phi_I=\theta-G_1(\theta)-\frac{\gamma}{r_\beta+r_\omega}(1-\theta)-\frac{r_\omega}{r_\omega+r_\beta}\big(1-G_1(\theta)\big),
\end{equation}

and finally using Eq.(\ref{Dtheta}) we can write a single differential
equation for $\theta$

\begin{equation}
  \dot{\theta}=-(r_\omega+r_\beta)\theta+\gamma(1-\theta)+r_\omega \big(1-G_1(\theta)\big).
\end{equation}

This equation together with $S=G_0(\theta)$,
$V=r_\omega/(r_w+r_\beta)\big(1-G_0(\theta)\big)$, $\dot{I}=\gamma R$
and $S+I+R+V=1$ describes the evolution of the fraction of
susceptible, infected, vaccinated and recovered individuals on a
complex network for continuous time.

Alternatively, we can derive Eq. (\ref{Eqtheta}) with respect to time
and use Eqs. (\ref{DPhiS}), (\ref{DPhiV}) and (\ref{DPhiR}) to write a
differential equation for $\Phi_I$

\begin{equation}
  \dot{\Phi}_I=-(r_\omega+r_\beta)\Phi_I+r_\beta G^{'}_1(\theta)\Phi_I -\gamma \Phi_I.
\end{equation}

This equation, along with Eqs. (\ref{Dtheta}), (\ref{DPhiS}),
(\ref{DPhiV}) are the continuous-time version of the set
(\ref{EqMiller}) of equations. For discrete time steps the derivatives
become forward finite differences, i.e., $f^{'}\big(x(t)\big)
\rightarrow f(x_{t+1}-x_{t})$, and also the rates $r_\beta$ and
$r_\omega$ become probabilities $\beta$ and $\omega$ respectively,
while $\gamma$ is replaced by the recovery time $t_r$.

\section{Temporal evolution of the discrete-time equations close to the threshold $V_L$ }

When iterating Eqs. (\ref{EqMiller}), one approach is to set
$\omega=0$ in the temporal step that $V(t)$ would surpass $V_L$,
ensuring that $V(t)\leq V_L$, but not that $V(t)=V_L$ at the steady
state. This would cause many fluctuations when computing $V(t)$ as a
function of $\beta$ at the steady state. Thus, to reproduce exactly
the results from the computational simulations another approach should
be used. Next we detail the procedure we use to avoid this
fluctuations. At time $n^\ast$ we calculate $\Delta V(n^\ast)$, and if
it turns out that $V(n^\ast)=V(n^\ast-1)+\Delta V(n^\ast)$ is greater
than $V_L$, then we calculate what is the value of $\omega^\ast$ that
satisfies $V(n^\ast)=V_L$. Thus, instead of setting $\omega=0$, we use
a smaller probability $\omega^\ast<\omega_0$, where $\omega_0$ is the
vaccination probability at the beginning of the process. This
adjustment of $\omega$ may have to be performed a couple of times
until finally $\omega=0$, but also $V(t)=V_L$.

Thus in order deal with the limit of the vaccine units while iterating
the equations, we have to use a vaccination probability that has a
slight dependence on time. Furthermore, we have to take into account
how this procedure affects $\Omega$, since it depends on $\omega$, as
we can see in Eq.~(\ref{Omegon}).
Suppose that we choose a set of parameters for which at some point the
population runs out of vaccines and we iterate the theoretical
equations. Then, at the beginning of the process $\Omega(t)$ is given
by Eq.~(\ref{Omegon}), and when the process ends
$\Omega=1-(1-\beta)^{t_r}$, which is simply the transmissibility of
the SIR model \cite{New_05}. Recall that $\Omega$ is the probability
that a node in state I infects one of its neighbors or induces its
vaccination during the time that this node remains in this
state. Consider that during the time that a node is infected the
probability of vaccination changes. Thus in this case the effective
probability of infection or immunization $\Omega$ is lower than the
one described in Eq.~(\ref{Omegon}) but higher than the
transmissibility of the SIR model. Considering the different
probabilities of vaccination that may have to be used during the
process we can write a general expression for $\Omega$ at time $t$,

\begin{equation}\label{Omegon2}
  \Omega_t=\sum_{n=1}^{t_r}(1-\beta)^{n-1} \bigg(\frac{w_{t-t_r+n}}{1-w_{t-t_r+n}}+\beta    \bigg)\prod_{j=1}^{n}\big(1-w_{t-t_r+j}\big).
\end{equation}

This expression takes in account that $\omega$ depends on time and is
proved in detail in Appendix C. If $\omega_t=\omega$ for all $t$, then
Eq. (\ref{Omegon2}) leads to Eq. (\ref{Omegon}):

\[
  \Omega=1-(1-\omega)^{t_r}(1-\beta)^{t_r}.
\]

\section{Derivation of $\Omega$ when $\omega$ depends on time}

$\Omega$ is the probability that a node infects one of its neighbors
or induces its immunization during the time that it is infected, which
is the recovery time $t_r$.  In the standard SIR model this
probability is known as the transmissibility $T$ and is calculated as
follows:

\begin{flalign}\label{Transs}
  T&=\beta+(1-\beta)\beta+(1-\beta)^2\beta+... \equiv \sum_{n=1}^{t_r}(1-\beta)^{n-1}\beta \\
   &= 1-(1-\beta)^{t_r}\nonumber
  \end{flalign}

At $n=1$ we simply consider the probability of infection $\beta$. At
$n=2$ we have to consider that the infection did not occur at $n=1$,
which happens with probability $1-\beta$. Next for $n=3$, now we have
to consider that there was no infection at $n=1$ and $n=2$, which
happens with probability $(1-\beta)^2$, and so on.

For the vaccination model we have to include the immunization
probability $\omega$. For simplicity we define $\Omega_1$ and
$\Omega_2$ as the effective probabilities of infection and
immunization respectively, during $t_r$ units of time. Thus, similar
to Eq.(\ref{Transs})

\begin{flalign}\label{Omeg_1_2}
  \Omega_1&= (1-\omega)\beta+(1-\omega)^2(1-\beta)\beta+(1-\omega)^3(1-\beta)^2\beta+...\\
  \Omega_2&= \omega+(1-\omega)(1-\beta)\omega+(1-\omega)^2(1-\beta)^2\omega+...  \nonumber
\end{flalign}

Note that if $\omega=0$, then $\Omega_1$ is the same as
Eq. (\ref{Transs}). Thus $\Omega_1$ plays the role of the
transmissibility in the dynamical vaccination model.
Furthermore, we can write Eq. (\ref{Omeg_1_2}) in a closed form

\begin{flalign}\label{Omeg_1_2_closed}
  \Omega_1&=\sum_{n=1}^{t_r-1}(1-\omega)^n(1-\beta)^{n-1}\beta=\frac{1-(1-\omega)^{t_r}(1-\beta)^{t_r}}{1-(1-\omega)(1-\beta)}(1-\omega)\beta\\
  \Omega_2&= \sum_{n=1}^{t_r-1}(1-\omega)^{n-1}(1-\beta)^{n-1}\omega=\frac{1-(1-\omega)^{t_r}(1-\beta)^{t_r}}{1-(1-\omega)(1-\beta)}\omega, \nonumber
\end{flalign}

and thus

\begin{equation}
  \Omega\equiv \Omega_1+\Omega_2= 1-(1-\omega)^{t_r}(1-\beta)^{t_r},
\end{equation}

which is Eq.~(\ref{Omegon}). When $\omega$ depends on time, a closed
expression can not be found. Suppose that $\omega=\omega_t$, and
consider a node that was infected at $t=0$, then Eq. (\ref{Omeg_1_2})
takes the form

\begin{flalign}
  \Omega_1= &(1-\omega_1)\beta+(1-\omega_1)(1-\omega_2)(1-\beta)\beta\\
  &+(1-\omega_1)(1-\omega_2)(1-\omega_3)(1-\beta)^2\beta+...\nonumber\\
  \Omega_2= &\omega_1+(1-\omega_1)(1-\beta)\omega_2+(1-\omega_1)(1-\omega_2)(1-\beta)^2\omega_2+..., \nonumber
  \end{flalign}

which can be summarized in the following expressions

\begin{flalign}\label{Omegon1_2_prod}
  \Omega_1&=\sum_{n=0}^{t_r}(1-\beta)^{n-1}\frac{\omega_n}{1-\omega_n}\prod_{j=1}^{n}(1-\omega_j) \\
  \Omega_2&=\sum_{n=0}^{t_r}(1-\beta)^{n-1}\beta\prod_{j=1}^{n}(1-\omega_j) . \nonumber
\end{flalign}

Eq. (\ref{Omegon1_2_prod}) represents the immunization and
infection transmission for $t=t_r$, since we are considering a single node
infected at $t=0$. We can generalize these equations for any time $t\geq t_r$

\begin{flalign}\label{Omegon1_2_prod2}
  \Omega_1(t)&=\sum_{n=0}^{t_r}(1-\beta)^{n-1}\frac{\omega_{t-t_r+n}}{1-\omega_{t-t_r+n}}\prod_{j=1}^{n}(1-\omega_{t-t_r+j})
  \\ \Omega_2(t)&=\sum_{n=0}^{t_r}(1-\beta)^{n-1}\beta
  \prod_{j=1}^{n}(1-\omega_{t-t_r+j}), \nonumber
\end{flalign}

and finally adding $\Omega_1(t)$ and $\Omega_1(t)$ leads to Eq. (\ref{Omegon2}).


\section{Computation of $\beta^*$ and $\beta^\dag$}

In the steady state of our model, the fraction of vaccinated
individuals when there is no limit in the number of vaccines is
\cite{Alvarez_2018}:
\begin{equation}
  V=C_\beta \big(1-G_0(1-T\;f_\infty)\big),
\end{equation}

where $C_\beta$ is the same factor used in Eq. (\ref{EqMiller}) and
$f_\infty$ satisfies the transcendental equation

\[
f_\infty=1-G_1(1-T\;f_\infty).
\]

$f_\infty$ is the probability that the branches of infection expand
indefinitely, and $T$ is the probability that an infected node spreads
the disease through a link, also known as
transmissibility\cite{Alvarez_2018}

\begin{equation}
T=\frac{1-(1-\omega)^{t_r}(1-\beta)^{t_r}}{\omega+\beta-\omega\beta}(1-\omega)\beta.  
  \end{equation}

Note that this expression is the same as $\Omega_1$ in
Eq. (\ref{Omeg_1_2_closed}).  If we compute the process for fixed $\omega$ and start
increasing $\beta$, when $\beta=\beta^\ast$ the fraction of vaccinated
nodes reach the limit $V_L$. Then for greater values of $\beta$ the
number of vaccinated individuals in the steady state is equal to
$V_L$, but beyond $\beta=\beta^\dag$ if exists, the vaccination
threshold is no more reached. If we observe closely Fig.~\ref{SRV_beta},
we see that for these values of $\beta$, the unlimited-vaccines
curve equals the vaccination limit $V_L$. Thus we can find
$\beta^\ast$ and $\beta^\dag$ by solving the following system:

\begin{flalign}\label{Vsystem}
  V_L&=1-G_0(1-T_{\beta^\ast,\omega}\;f_\infty),\\
  f_\infty&=1-G_1(1-T_{\beta^\ast,\omega}\;f_\infty),\nonumber
  \end{flalign}

where $T_{\beta^\ast,\omega}$ is the transmissibility for
$\beta=\beta^\ast$ and fixed $\omega$. The same applies for $\beta^{\dag}$ and
$T_{\beta^\dag,\omega}$.
For a Poisson network $G_0(y) \equiv G_1(y)$ and
hence we can write a single transcendental equation to find $\beta^\ast$
or $\beta^{\dag}$ :

\begin{equation}\label{VTrascendental1}
  x=\frac{\omega}{V_L(1-\omega)}\Bigg\{1-exp\bigg[-\langle k \rangle T_{x,\omega}V_L\frac{\omega+(1-\omega)x}{\omega}\bigg]\Bigg\}-\frac{\omega}{1-\omega}.
\end{equation}

This equation has one, two, or none solutions between $0$ and $1$
depending on the parameters. Two different solutions correspond to
$\beta^\ast$ and $\beta^{\dag}$ while a single solution means that
$\beta^{\dag}$ does not exist. Moreover if $V_L$ is large enough,
there is no solution, which means that there are always available
vaccines when needed.

Based on a similar reasoning we can find $w^\ast$ and $w^\dag$ for
fixed $\beta$. From Eqs.~(\ref{Vsystem}) we can derive a similar expression to Eq.~(\ref{VTrascendental1}) for a Poisson network:

\begin{equation}\label{VTrascendental2}
  y=\frac{V_L(1-y)\beta}{1-exp\bigg[-\langle k \rangle T_{\beta,y}V_L\frac{y+(1-y)\beta}{\omega}\bigg]-V_L},
\end{equation}

which has two or none solutions depending on $V_L$.

\begin{figure}[h]
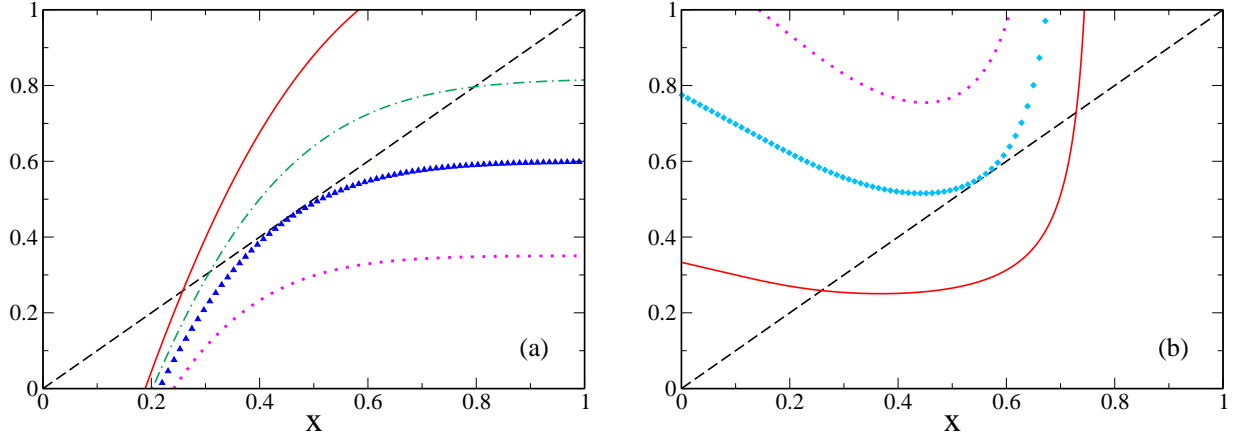

  \begin{center}
     \includegraphics[width=0.47\textwidth]{Graphsol1.eps}
      \hspace{0.5cm}
     \includegraphics[width=0.47\textwidth]{Graphsol2.eps}  
  \end{center}
  \caption{Graphical solution of Eqs.~(\ref{VTrascendental1}) and
    (\ref{VTrascendental2}). The solutions are given by the
    intersection of the r.h.s. of the equations with the identity
    (\protect\blackdashed). (a) Solutions of
    Eq.~(\ref{VTrascendental1}) for $\omega=0.45$. Solutions of
    Eq.~(\ref{VTrascendental2}) for $\beta=0.5$. The curves
    represent different values of vaccination limits: $V_L=0.4$
    (\protect\redsolid), $V_L=0.5$ (\protect\verdecitodashdotted),
    $V_L=0.574$ (\protect\bluetriangle), $V_L=0.606$
    (\protect\skydiamondII) and $V_L=0.7$
    (\protect\magdotted).}\label{Graphsol}
  \end{figure}

In Fig.~\ref{Graphsol} we show the graphical solution of the previous
equations for different values of $V_L$. In (a) we graphically solve
Eq.~(\ref{VTrascendental1}) for fixed $\omega=0.45$. For $V_L=0.4$ the
curve intersects the identity in only one point, which corresponds to
$\beta^\ast$. On the contrary for $V_L=0.5$ there are two intersection
points, which denotes the existence of $\beta^\dag$. For $V_L=0.574$
the curve is tangential to the identity, and thus
$\beta^\ast=\beta^\dag$. When $V_L>0.574$ there is no
solution. Similarly we show in Fig.~\ref{Graphsol} (b) the solutions
of Eq.~(\ref{VTrascendental1}) for fixed $\beta=0.5$. In this case for
$V_L=0.4$ now we have two intersection points, which correspond to
$\omega^\ast$ and $\omega^\dag$. For $V_L=0.606$ the curve is
tangential to the identity and hence there is a single solution, which
means $\omega^\ast=\omega^\dag$. As we can see beyond this point there
is no solution.

\vspace{10cm}

\section{Supplementary figures of the steady state}

\begin{figure}[H]
  \begin{center}
     \includegraphics[width=0.47\textwidth]{tr3_V_w_045_ER_km_10.eps}
      \hspace{0.5cm}
      \includegraphics[width=0.47\textwidth]{tr3_R_w_045_ER_km_10.eps}
      \vspace{0.5cm}
      \includegraphics[width=0.47\textwidth]{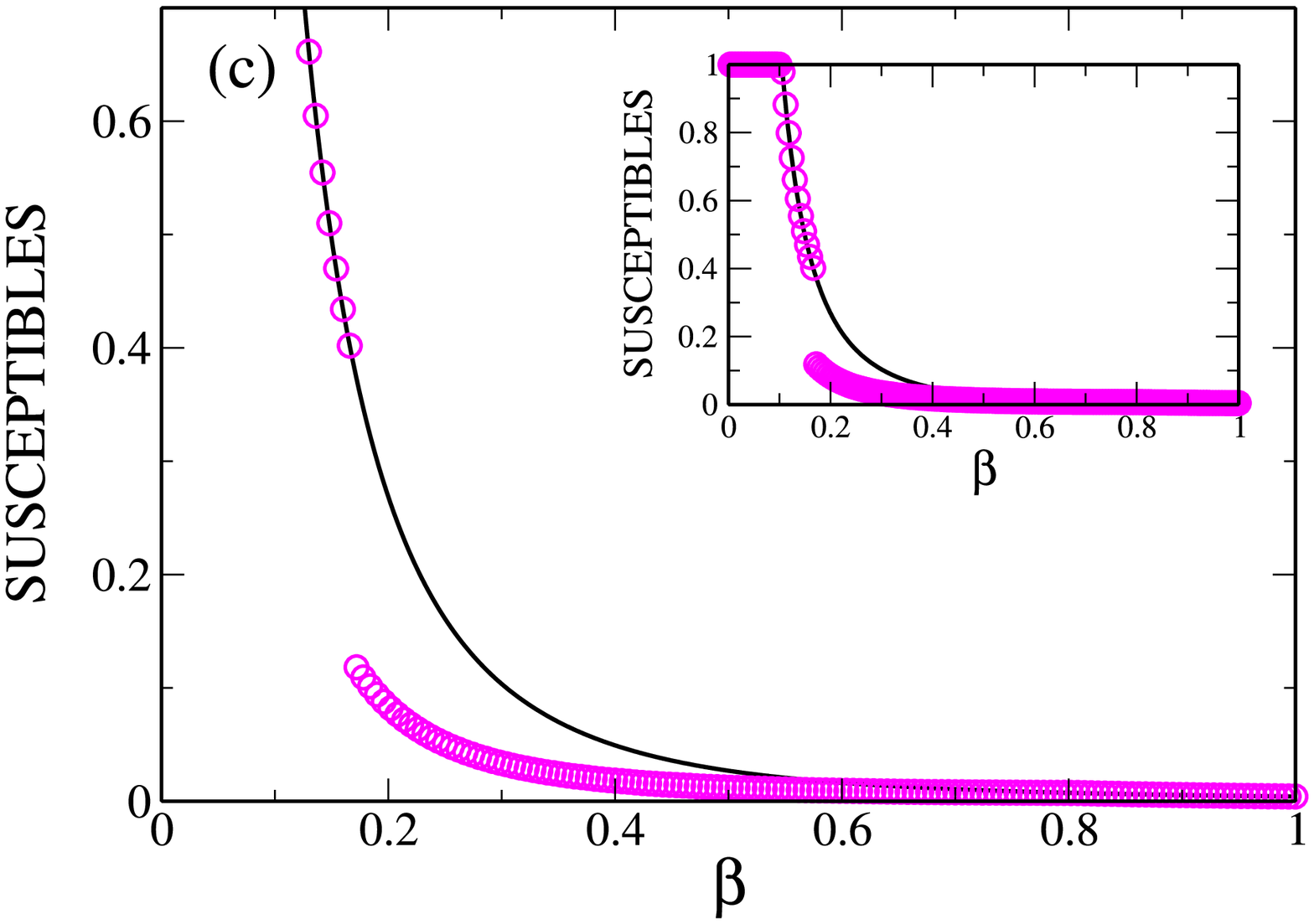}
     \hspace{0.5cm}
     \includegraphics[width=0.47\textwidth]{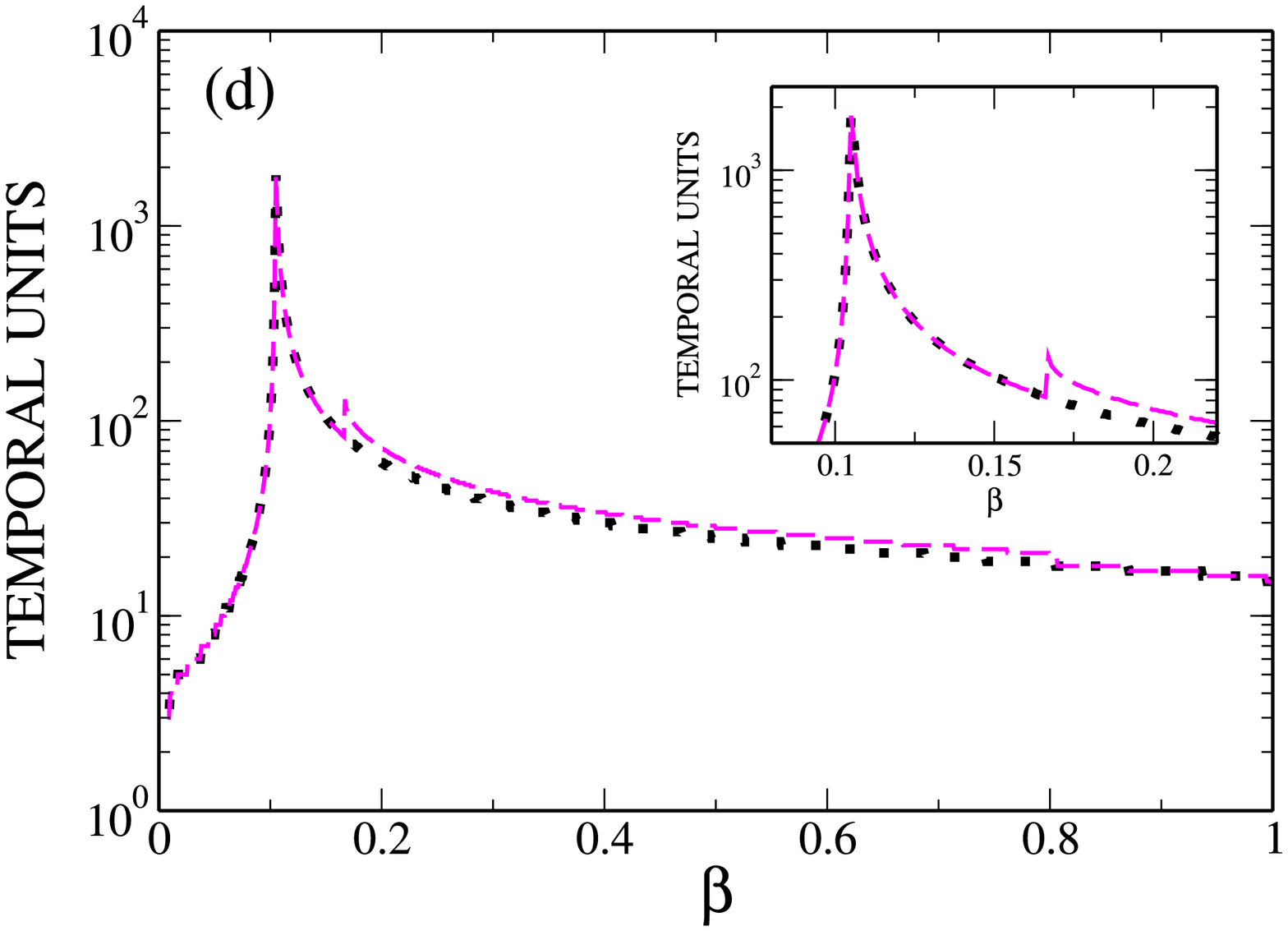}
  \end{center}
  \caption{Fraction of vaccinated, recovered and susceptible
    individuals at the steady state as a function of the infection
    probability $\beta$. The degree distribution is ER with
    $\langle k \rangle=10$ $k_{min}=0$ and $k_{max}=40$, and the
    recovery time $t_r=3$. The vaccination probability is
    $\omega=0.45$ and the vaccination limits $V_L=1$
    (\protect\blacksolid) and $V_L=0.5$ (\protect\magcirc). The vertical
    dashed lines indicate the values of $\beta^\ast$ and $\beta^\dag$
    for $V_L=0.5$. In (c), we show in the inset the
    full curve of susceptible nodes. In the last figure we compute theoretically the
    time it take to the process to reach the steady state. Since the
    peaks are very close we show an inset for a better
    visualization. In this figure $V_L=1$ (\protect\blackdotted),
    $0.5$ (\protect\magdashed).}\label{SRV_beta_tr3}
  \end{figure}

In Fig.~\ref{SRV_beta_tr3} we show how the steady state changes when
the recovery time $t_r$ is greater than $1$. Note that this figure is similar to
Fig.~\ref{SRV_beta}, with the same vaccination probability
$\omega=0.45$, vaccination limits $V_L=1$ and $V_L=0.5$, and a
different recovery time $t_r=3$. We see that $\beta^*$, unlike
$\beta^\dag$ which barely changes, is lower than for
$t_r=1$. Furthermore, because the individuals are infected during a
larger period of time, the fraction of recovered is larger and so is
the discontinuous jump.

\begin{figure}[H]
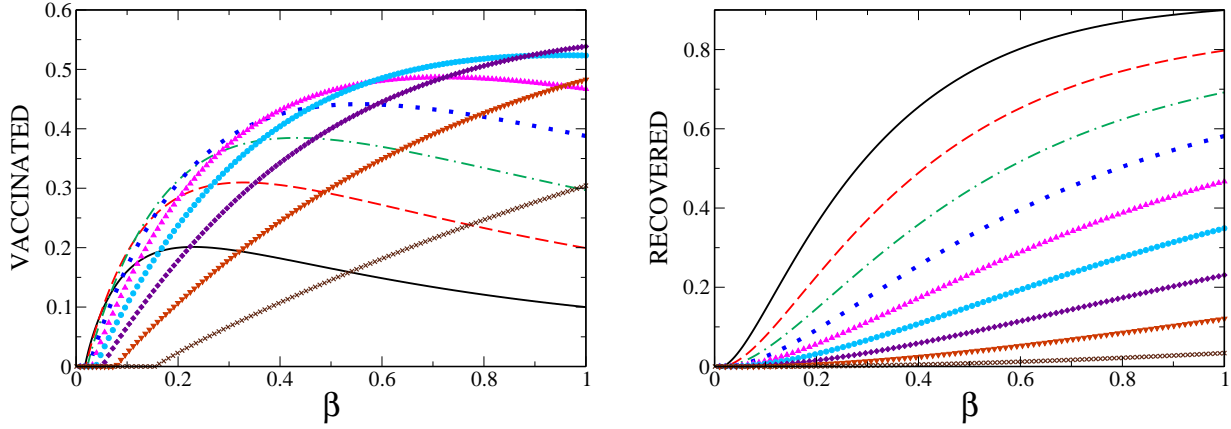

  \begin{center}
     \includegraphics[width=0.47\textwidth]{V_SF_w.eps}
      \hspace{0.5cm}
     \includegraphics[width=0.47\textwidth]{R_SF_w.eps}  
     \end{center}
  \caption{Fraction of vaccinated and recovered individuals at the
    steady state as a function of the infection probability $\beta$
    for $t_r=1$ and $V_L=1$. The degree distribution is power law or
    scale-free with exponent $\lambda=2.2$, $k_{min}=3$ and
    $k_{max}=500$, and the recovery time $t_r=3$. The different curves
    represent different vaccination probabilities, $\omega=0.1$
    (\protect\blacksolid), $\omega=0.2$ (\protect\reddashed),
    $\omega=0.3$ (\protect\verdecitodashdot), $\omega=0.4$
    (\protect\bluedotted), $\omega=0.5$ (\protect\magtrg),
    $\omega=0.6$ (\protect\skycirc), $\omega=0.7$
    (\protect\violdiamond), $\omega=0.8$ (\protect\orgIItrginv) and
    $\omega=0.9$ (\protect\sepiacross)}\label{Vac_SF}
  \end{figure}

On the other hand in Fig.~\ref{Vac_SF} (a) we show the fracion of
vaccinated and recovered individuals in the steady state as a function
of the infection probability $\beta$, for a network with a power law
degree distribution, and for different vaccination probabilities
$\omega$. The vaccination limit and the recovery time are fixed,
$V_L=1$ and $t_r=1$. We observe that for low values of $\omega$ the
curves of vaccinated individuals exhibit a maximum while, for larger values they are monotonically increasing. As
explained in the main text, $\beta^\dag$ exists as long as the curve of
vaccinated individuals has a maximum, hence in this case for
$\omega>0.6$ only $\beta^*$ exists. In addition, in Fig.~\ref{Vac_SF}
(b) we show the fraction of recovered individuals, which as expected
decrease when the vaccination probability is higher.

\begin{thebibliography}{55}%
\makeatletter
\providecommand \@ifxundefined [1]{%
 \@ifx{#1\undefined}
}%
\providecommand \@ifnum [1]{%
 \ifnum #1\expandafter \@firstoftwo
 \else \expandafter \@secondoftwo
 \fi
}%
\providecommand \@ifx [1]{%
 \ifx #1\expandafter \@firstoftwo
 \else \expandafter \@secondoftwo
 \fi
}%
\providecommand \natexlab [1]{#1}%
\providecommand \enquote  [1]{``#1''}%
\providecommand \bibnamefont  [1]{#1}%
\providecommand \bibfnamefont [1]{#1}%
\providecommand \citenamefont [1]{#1}%
\providecommand \href@noop [0]{\@secondoftwo}%
\providecommand \href [0]{\begingroup \@sanitize@url \@href}%
\providecommand \@href[1]{\@@startlink{#1}\@@href}%
\providecommand \@@href[1]{\endgroup#1\@@endlink}%
\providecommand \@sanitize@url [0]{\catcode `\\12\catcode `\$12\catcode
  `\&12\catcode `\#12\catcode `\^12\catcode `\_12\catcode `\%12\relax}%
\providecommand \@@startlink[1]{}%
\providecommand \@@endlink[0]{}%
\providecommand \url  [0]{\begingroup\@sanitize@url \@url }%
\providecommand \@url [1]{\endgroup\@href {#1}{\urlprefix }}%
\providecommand \urlprefix  [0]{URL }%
\providecommand \Eprint [0]{\href }%
\providecommand \doibase [0]{http://dx.doi.org/}%
\providecommand \selectlanguage [0]{\@gobble}%
\providecommand \bibinfo  [0]{\@secondoftwo}%
\providecommand \bibfield  [0]{\@secondoftwo}%
\providecommand \translation [1]{[#1]}%
\providecommand \BibitemOpen [0]{}%
\providecommand \bibitemStop [0]{}%
\providecommand \bibitemNoStop [0]{.\EOS\space}%
\providecommand \EOS [0]{\spacefactor3000\relax}%
\providecommand \BibitemShut  [1]{\csname bibitem#1\endcsname}%
\let\auto@bib@innerbib\@empty
\bibitem [{\citenamefont {Boccaletti}\ \emph {et~al.}(2006)\citenamefont
  {Boccaletti}, \citenamefont {Latora}, \citenamefont {Moreno}, \citenamefont
  {Chavez},\ and\ \citenamefont {Hwang}}]{Boc_01}%
  \BibitemOpen
  \bibfield  {author} {\bibinfo {author} {\bibfnamefont {S.}~\bibnamefont
  {Boccaletti}}, \bibinfo {author} {\bibfnamefont {V.}~\bibnamefont {Latora}},
  \bibinfo {author} {\bibfnamefont {Y.}~\bibnamefont {Moreno}}, \bibinfo
  {author} {\bibfnamefont {M.}~\bibnamefont {Chavez}}, \ and\ \bibinfo {author}
  {\bibfnamefont {D.}~\bibnamefont {Hwang}},\ }\href@noop {} {\bibfield
  {journal} {\bibinfo  {journal} {Phys. Rep.}\ }\textbf {\bibinfo {volume}
  {424}},\ \bibinfo {pages} {175} (\bibinfo {year} {2006})}\BibitemShut
  {NoStop}%
\bibitem [{\citenamefont {Barrat}\ \emph {et~al.}(2004)\citenamefont {Barrat},
  \citenamefont {Barth{\'e}lemy}, \citenamefont {Pastor-Satorras},\ and\
  \citenamefont {Vespignani}}]{barrat_04}%
  \BibitemOpen
  \bibfield  {author} {\bibinfo {author} {\bibfnamefont {A.}~\bibnamefont
  {Barrat}}, \bibinfo {author} {\bibfnamefont {M.}~\bibnamefont
  {Barth{\'e}lemy}}, \bibinfo {author} {\bibfnamefont {R.}~\bibnamefont
  {Pastor-Satorras}}, \ and\ \bibinfo {author} {\bibfnamefont {A.}~\bibnamefont
  {Vespignani}},\ }\href@noop {} {\bibfield  {journal} {\bibinfo  {journal}
  {Proc. Natl. Acad. Sci. USA}\ }\textbf {\bibinfo {volume} {101}},\ \bibinfo
  {pages} {3747} (\bibinfo {year} {2004})}\BibitemShut {NoStop}%
\bibitem [{\citenamefont {Newman}(2010)}]{New_10}%
  \BibitemOpen
  \bibfield  {author} {\bibinfo {author} {\bibfnamefont {M.~E.~J.}\
  \bibnamefont {Newman}},\ }\href@noop {} {\emph {\bibinfo {title} {Networks:
  An Introduction}}}\ (\bibinfo  {publisher} {Oxford University Press},\
  \bibinfo {year} {2010})\BibitemShut {NoStop}%
\bibitem [{\citenamefont {Cohen}\ and\ \citenamefont {Havlin}(2010)}]{Coh_10}%
  \BibitemOpen
  \bibfield  {author} {\bibinfo {author} {\bibfnamefont {R.}~\bibnamefont
  {Cohen}}\ and\ \bibinfo {author} {\bibfnamefont {S.}~\bibnamefont {Havlin}},\
  }\href@noop {} {\emph {\bibinfo {title} {Complex Networks: Structure,
  Robustness and Function}}}\ (\bibinfo  {publisher} {Cambridge University
  Press},\ \bibinfo {year} {2010})\BibitemShut {NoStop}%
\bibitem [{\citenamefont {Cattuto}\ \emph {et~al.}(2010)\citenamefont
  {Cattuto}, \citenamefont {den Broeck}, \citenamefont {Barrat}, \citenamefont
  {Colizza}, \citenamefont {Pinton},\ and\ \citenamefont
  {Vespignani}}]{Catt_01}%
  \BibitemOpen
  \bibfield  {author} {\bibinfo {author} {\bibfnamefont {C.}~\bibnamefont
  {Cattuto}}, \bibinfo {author} {\bibfnamefont {W.~V.}\ \bibnamefont {den
  Broeck}}, \bibinfo {author} {\bibfnamefont {A.}~\bibnamefont {Barrat}},
  \bibinfo {author} {\bibfnamefont {V.}~\bibnamefont {Colizza}}, \bibinfo
  {author} {\bibfnamefont {J.-F.}\ \bibnamefont {Pinton}}, \ and\ \bibinfo
  {author} {\bibfnamefont {A.}~\bibnamefont {Vespignani}},\ }\href@noop {}
  {\bibfield  {journal} {\bibinfo  {journal} {PLoS ONE}\ }\textbf {\bibinfo
  {volume} {5}},\ \bibinfo {pages} {e11596} (\bibinfo {year}
  {2010})}\BibitemShut {NoStop}%
\bibitem [{\citenamefont {Gonzalez}\ \emph {et~al.}(2008)\citenamefont
  {Gonzalez}, \citenamefont {Hidalgo},\ and\ \citenamefont
  {Barabasi}}]{gonzalez_08}%
  \BibitemOpen
  \bibfield  {author} {\bibinfo {author} {\bibfnamefont {M.~C.}\ \bibnamefont
  {Gonzalez}}, \bibinfo {author} {\bibfnamefont {C.~A.}\ \bibnamefont
  {Hidalgo}}, \ and\ \bibinfo {author} {\bibfnamefont {A.-L.}\ \bibnamefont
  {Barabasi}},\ }\href@noop {} {\bibfield  {journal} {\bibinfo  {journal}
  {Nature}\ }\textbf {\bibinfo {volume} {453}},\ \bibinfo {pages} {779}
  (\bibinfo {year} {2008})}\BibitemShut {NoStop}%
\bibitem [{\citenamefont {G{\'o}mez-Garde{\~n}es}\ \emph
  {et~al.}(2008)\citenamefont {G{\'o}mez-Garde{\~n}es}, \citenamefont {Latora},
  \citenamefont {Moreno},\ and\ \citenamefont {Profumo}}]{gardenes_08}%
  \BibitemOpen
  \bibfield  {author} {\bibinfo {author} {\bibfnamefont {J.}~\bibnamefont
  {G{\'o}mez-Garde{\~n}es}}, \bibinfo {author} {\bibfnamefont {V.}~\bibnamefont
  {Latora}}, \bibinfo {author} {\bibfnamefont {Y.}~\bibnamefont {Moreno}}, \
  and\ \bibinfo {author} {\bibfnamefont {E.}~\bibnamefont {Profumo}},\
  }\href@noop {} {\bibfield  {journal} {\bibinfo  {journal} {Proceedings of the
  National Academy of Sciences}\ }\textbf {\bibinfo {volume} {105}},\ \bibinfo
  {pages} {1399} (\bibinfo {year} {2008})}\BibitemShut {NoStop}%
\bibitem [{\citenamefont {Volz}\ \emph {et~al.}(2011)\citenamefont {Volz},
  \citenamefont {Miller}, \citenamefont {Galvani},\ and\ \citenamefont
  {Meyers}}]{volz_11}%
  \BibitemOpen
  \bibfield  {author} {\bibinfo {author} {\bibfnamefont {E.~M.}\ \bibnamefont
  {Volz}}, \bibinfo {author} {\bibfnamefont {J.~C.}\ \bibnamefont {Miller}},
  \bibinfo {author} {\bibfnamefont {A.}~\bibnamefont {Galvani}}, \ and\
  \bibinfo {author} {\bibfnamefont {L.~A.}\ \bibnamefont {Meyers}},\
  }\href@noop {} {\bibfield  {journal} {\bibinfo  {journal} {PLoS Comput.
  Biol.}\ }\textbf {\bibinfo {volume} {7}},\ \bibinfo {pages} {e1002042}
  (\bibinfo {year} {2011})}\BibitemShut {NoStop}%
\bibitem [{\citenamefont {Bailey}(1975)}]{Bailey_75}%
  \BibitemOpen
  \bibfield  {author} {\bibinfo {author} {\bibfnamefont {N.~T.~J.}\
  \bibnamefont {Bailey}},\ }\href@noop {} {\emph {\bibinfo {title} {{The
  Mathematical Theory of Infectious Diseases}}}}\ (\bibinfo  {publisher}
  {Griffin, London},\ \bibinfo {year} {1975})\BibitemShut {NoStop}%
\bibitem [{\citenamefont {Anderson}\ and\ \citenamefont
  {May}(1992)}]{Ander_91}%
  \BibitemOpen
  \bibfield  {author} {\bibinfo {author} {\bibfnamefont {R.~M.}\ \bibnamefont
  {Anderson}}\ and\ \bibinfo {author} {\bibfnamefont {R.~M.}\ \bibnamefont
  {May}},\ }\href@noop {} {\emph {\bibinfo {title} {{Infectious Diseases of
  Humans: Dynamics and Control}}}}\ (\bibinfo  {publisher} {Oxford University
  Press, Oxford},\ \bibinfo {year} {1992})\BibitemShut {NoStop}%
\bibitem [{\citenamefont {Pastor-Satorras}\ and\ \citenamefont
  {Vespignani}(2002)}]{past_03}%
  \BibitemOpen
  \bibfield  {author} {\bibinfo {author} {\bibfnamefont {R.}~\bibnamefont
  {Pastor-Satorras}}\ and\ \bibinfo {author} {\bibfnamefont {A.}~\bibnamefont
  {Vespignani}},\ }\href@noop {} {\bibfield  {journal} {\bibinfo  {journal}
  {Phys. Rev. E}\ }\textbf {\bibinfo {volume} {65}},\ \bibinfo {pages} {036104}
  (\bibinfo {year} {2002})}\BibitemShut {NoStop}%
\bibitem [{\citenamefont {De~Domenico}\ \emph {et~al.}(2016)\citenamefont
  {De~Domenico}, \citenamefont {Granell}, \citenamefont {Porter},\ and\
  \citenamefont {Arenas}}]{Arenas_16}%
  \BibitemOpen
  \bibfield  {author} {\bibinfo {author} {\bibfnamefont {M.}~\bibnamefont
  {De~Domenico}}, \bibinfo {author} {\bibfnamefont {C.}~\bibnamefont
  {Granell}}, \bibinfo {author} {\bibfnamefont {M.~A.}\ \bibnamefont {Porter}},
  \ and\ \bibinfo {author} {\bibfnamefont {A.}~\bibnamefont {Arenas}},\
  }\href@noop {} {\bibfield  {journal} {\bibinfo  {journal} {Nature Physics}\
  }\textbf {\bibinfo {volume} {12}} (\bibinfo {year} {2016})}\BibitemShut
  {NoStop}%
\bibitem [{\citenamefont {Castellano}\ and\ \citenamefont
  {Pastor-Satorras}(2010)}]{castellano_10}%
  \BibitemOpen
  \bibfield  {author} {\bibinfo {author} {\bibfnamefont {C.}~\bibnamefont
  {Castellano}}\ and\ \bibinfo {author} {\bibfnamefont {R.}~\bibnamefont
  {Pastor-Satorras}},\ }\href@noop {} {\bibfield  {journal} {\bibinfo
  {journal} {Phys. Rev. Lett}\ }\textbf {\bibinfo {volume} {105}},\ \bibinfo
  {pages} {218701} (\bibinfo {year} {2010})}\BibitemShut {NoStop}%
\bibitem [{\citenamefont {Pastor-Satorras}\ \emph {et~al.}(2015)\citenamefont
  {Pastor-Satorras}, \citenamefont {Castellano}, \citenamefont {Van~Mieghem},\
  and\ \citenamefont {Vespignani}}]{Pastor_15}%
  \BibitemOpen
  \bibfield  {author} {\bibinfo {author} {\bibfnamefont {R.}~\bibnamefont
  {Pastor-Satorras}}, \bibinfo {author} {\bibfnamefont {C.}~\bibnamefont
  {Castellano}}, \bibinfo {author} {\bibfnamefont {P.}~\bibnamefont
  {Van~Mieghem}}, \ and\ \bibinfo {author} {\bibfnamefont {A.}~\bibnamefont
  {Vespignani}},\ }\href@noop {} {\bibfield  {journal} {\bibinfo  {journal}
  {Rev. Mod. Phys.}\ }\textbf {\bibinfo {volume} {87}},\ \bibinfo {pages} {925}
  (\bibinfo {year} {2015})}\BibitemShut {NoStop}%
\bibitem [{\citenamefont {Wang}\ \emph {et~al.}(2017)\citenamefont {Wang},
  \citenamefont {Tang}, \citenamefont {Stanley},\ and\ \citenamefont
  {Braunstein}}]{Braunstein_16}%
  \BibitemOpen
  \bibfield  {author} {\bibinfo {author} {\bibfnamefont {W.}~\bibnamefont
  {Wang}}, \bibinfo {author} {\bibfnamefont {M.}~\bibnamefont {Tang}}, \bibinfo
  {author} {\bibfnamefont {H.~E.}\ \bibnamefont {Stanley}}, \ and\ \bibinfo
  {author} {\bibfnamefont {L.~A.}\ \bibnamefont {Braunstein}},\ }\href@noop {}
  {\bibfield  {journal} {\bibinfo  {journal} {Reports on Progress in Physics}\
  }\textbf {\bibinfo {volume} {80}},\ \bibinfo {pages} {036603} (\bibinfo
  {year} {2017})}\BibitemShut {NoStop}%
\bibitem [{\citenamefont {Cohen}\ \emph {et~al.}(2003)\citenamefont {Cohen},
  \citenamefont {Havlin},\ and\ \citenamefont {{ben-Avraham}}}]{Coh_03}%
  \BibitemOpen
  \bibfield  {author} {\bibinfo {author} {\bibfnamefont {R.}~\bibnamefont
  {Cohen}}, \bibinfo {author} {\bibfnamefont {S.}~\bibnamefont {Havlin}}, \
  and\ \bibinfo {author} {\bibfnamefont {D.}~\bibnamefont {{ben-Avraham}}},\
  }\href@noop {} {\bibfield  {journal} {\bibinfo  {journal} {Phys. Rev. Lett.}\
  }\textbf {\bibinfo {volume} {91}},\ \bibinfo {pages} {247901} (\bibinfo
  {year} {2003})}\BibitemShut {NoStop}%
\bibitem [{\citenamefont {Gallos}\ \emph {et~al.}(2007)\citenamefont {Gallos},
  \citenamefont {Liljeros}, \citenamefont {Argyrakis}, \citenamefont {Bunde},\
  and\ \citenamefont {Havlin}}]{Gallos_07}%
  \BibitemOpen
  \bibfield  {author} {\bibinfo {author} {\bibfnamefont {L.~K.}\ \bibnamefont
  {Gallos}}, \bibinfo {author} {\bibfnamefont {F.}~\bibnamefont {Liljeros}},
  \bibinfo {author} {\bibfnamefont {P.}~\bibnamefont {Argyrakis}}, \bibinfo
  {author} {\bibfnamefont {A.}~\bibnamefont {Bunde}}, \ and\ \bibinfo {author}
  {\bibfnamefont {S.}~\bibnamefont {Havlin}},\ }\href@noop {} {\bibfield
  {journal} {\bibinfo  {journal} {Phys. Rev. E}\ }\textbf {\bibinfo {volume}
  {75}},\ \bibinfo {pages} {045104} (\bibinfo {year} {2007})}\BibitemShut
  {NoStop}%
\bibitem [{\citenamefont {Wang}\ \emph {et~al.}(2016)\citenamefont {Wang},
  \citenamefont {Bauch}, \citenamefont {Bhattacharyya}, \citenamefont
  {d'Onofrio}, \citenamefont {Manfredi}, \citenamefont {Perc}, \citenamefont
  {Perra}, \citenamefont {Salath{\'e}},\ and\ \citenamefont {Zhao}}]{wang_16}%
  \BibitemOpen
  \bibfield  {author} {\bibinfo {author} {\bibfnamefont {Z.}~\bibnamefont
  {Wang}}, \bibinfo {author} {\bibfnamefont {C.~T.}\ \bibnamefont {Bauch}},
  \bibinfo {author} {\bibfnamefont {S.}~\bibnamefont {Bhattacharyya}}, \bibinfo
  {author} {\bibfnamefont {A.}~\bibnamefont {d'Onofrio}}, \bibinfo {author}
  {\bibfnamefont {P.}~\bibnamefont {Manfredi}}, \bibinfo {author}
  {\bibfnamefont {M.}~\bibnamefont {Perc}}, \bibinfo {author} {\bibfnamefont
  {N.}~\bibnamefont {Perra}}, \bibinfo {author} {\bibfnamefont
  {M.}~\bibnamefont {Salath{\'e}}}, \ and\ \bibinfo {author} {\bibfnamefont
  {D.}~\bibnamefont {Zhao}},\ }\href@noop {} {\bibfield  {journal} {\bibinfo
  {journal} {Physics Reports}\ }\textbf {\bibinfo {volume} {664}},\ \bibinfo
  {pages} {1} (\bibinfo {year} {2016})}\BibitemShut {NoStop}%
\bibitem [{\citenamefont {Valdez}\ \emph {et~al.}(2013)\citenamefont {Valdez},
  \citenamefont {Macri},\ and\ \citenamefont {Braunstein}}]{Valdez_13}%
  \BibitemOpen
  \bibfield  {author} {\bibinfo {author} {\bibfnamefont {L.~D.}\ \bibnamefont
  {Valdez}}, \bibinfo {author} {\bibfnamefont {P.~A.}\ \bibnamefont {Macri}}, \
  and\ \bibinfo {author} {\bibfnamefont {L.~A.}\ \bibnamefont {Braunstein}},\
  }\href@noop {} {\bibfield  {journal} {\bibinfo  {journal} {Physica A:
  Statistical Mechanics and its Applications}\ }\textbf {\bibinfo {volume}
  {392}},\ \bibinfo {pages} {4172 } (\bibinfo {year} {2013})}\BibitemShut
  {NoStop}%
\bibitem [{\citenamefont {Wang}\ \emph {et~al.}(2012)\citenamefont {Wang},
  \citenamefont {Zhang}, \citenamefont {Huang},\ and\ \citenamefont
  {Li}}]{wang_12}%
  \BibitemOpen
  \bibfield  {author} {\bibinfo {author} {\bibfnamefont {L.}~\bibnamefont
  {Wang}}, \bibinfo {author} {\bibfnamefont {Y.}~\bibnamefont {Zhang}},
  \bibinfo {author} {\bibfnamefont {T.}~\bibnamefont {Huang}}, \ and\ \bibinfo
  {author} {\bibfnamefont {X.}~\bibnamefont {Li}},\ }\href@noop {} {\bibfield
  {journal} {\bibinfo  {journal} {Phys. Rev. E}\ }\textbf {\bibinfo {volume}
  {86}},\ \bibinfo {pages} {032901} (\bibinfo {year} {2012})}\BibitemShut
  {NoStop}%
\bibitem [{\citenamefont {Newman}(2002)}]{New_05}%
  \BibitemOpen
  \bibfield  {author} {\bibinfo {author} {\bibfnamefont {M.~E.~J.}\
  \bibnamefont {Newman}},\ }\href@noop {} {\bibfield  {journal} {\bibinfo
  {journal} {Phys. Rev.E}\ }\textbf {\bibinfo {volume} {66}},\ \bibinfo {pages}
  {016128} (\bibinfo {year} {2002})}\BibitemShut {NoStop}%
\bibitem [{\citenamefont {Moreno}\ \emph {et~al.}(2002)\citenamefont {Moreno},
  \citenamefont {Pastor-Satorras},\ and\ \citenamefont
  {Vespignani}}]{moreno_02}%
  \BibitemOpen
  \bibfield  {author} {\bibinfo {author} {\bibfnamefont {Y.}~\bibnamefont
  {Moreno}}, \bibinfo {author} {\bibfnamefont {R.}~\bibnamefont
  {Pastor-Satorras}}, \ and\ \bibinfo {author} {\bibfnamefont {A.}~\bibnamefont
  {Vespignani}},\ }\href@noop {} {\bibfield  {journal} {\bibinfo  {journal}
  {The European Physical Journal B-Condensed Matter and Complex Systems}\
  }\textbf {\bibinfo {volume} {26}},\ \bibinfo {pages} {521} (\bibinfo {year}
  {2002})}\BibitemShut {NoStop}%
\bibitem [{\citenamefont {Volz}(2008)}]{volzsir}%
  \BibitemOpen
  \bibfield  {author} {\bibinfo {author} {\bibfnamefont {E.}~\bibnamefont
  {Volz}},\ }\href@noop {} {\bibfield  {journal} {\bibinfo  {journal} {Journal
  of mathematical biology}\ }\textbf {\bibinfo {volume} {56}},\ \bibinfo
  {pages} {293} (\bibinfo {year} {2008})}\BibitemShut {NoStop}%
\bibitem [{\citenamefont {Miller}(2011)}]{miller_11}%
  \BibitemOpen
  \bibfield  {author} {\bibinfo {author} {\bibfnamefont {J.~C.}\ \bibnamefont
  {Miller}},\ }\href@noop {} {\bibfield  {journal} {\bibinfo  {journal}
  {Journal of mathematical biology}\ }\textbf {\bibinfo {volume} {62}},\
  \bibinfo {pages} {349} (\bibinfo {year} {2011})}\BibitemShut {NoStop}%
\bibitem [{\citenamefont {Colizza}\ \emph {et~al.}(2006)\citenamefont
  {Colizza}, \citenamefont {Barrat}, \citenamefont {Barth{\'e}lemy},\ and\
  \citenamefont {Vespignani}}]{Colizza_06}%
  \BibitemOpen
  \bibfield  {author} {\bibinfo {author} {\bibfnamefont {V.}~\bibnamefont
  {Colizza}}, \bibinfo {author} {\bibfnamefont {A.}~\bibnamefont {Barrat}},
  \bibinfo {author} {\bibfnamefont {M.}~\bibnamefont {Barth{\'e}lemy}}, \ and\
  \bibinfo {author} {\bibfnamefont {A.}~\bibnamefont {Vespignani}},\
  }\href@noop {} {\bibfield  {journal} {\bibinfo  {journal} {Proceedings of the
  National Academy of Sciences}\ }\textbf {\bibinfo {volume} {103}},\ \bibinfo
  {pages} {2015} (\bibinfo {year} {2006})}\BibitemShut {NoStop}%
\bibitem [{\citenamefont {Meyers}\ \emph {et~al.}(2005)\citenamefont {Meyers},
  \citenamefont {Pourbohloul}, \citenamefont {Newman}, \citenamefont
  {Skowronski},\ and\ \citenamefont {Brunham}}]{meyers_sars}%
  \BibitemOpen
  \bibfield  {author} {\bibinfo {author} {\bibfnamefont {L.~A.}\ \bibnamefont
  {Meyers}}, \bibinfo {author} {\bibfnamefont {B.}~\bibnamefont {Pourbohloul}},
  \bibinfo {author} {\bibfnamefont {M.~E.}\ \bibnamefont {Newman}}, \bibinfo
  {author} {\bibfnamefont {D.~M.}\ \bibnamefont {Skowronski}}, \ and\ \bibinfo
  {author} {\bibfnamefont {R.~C.}\ \bibnamefont {Brunham}},\ }\href@noop {}
  {\bibfield  {journal} {\bibinfo  {journal} {Journal of theoretical biology}\
  }\textbf {\bibinfo {volume} {232}},\ \bibinfo {pages} {71} (\bibinfo {year}
  {2005})}\BibitemShut {NoStop}%
\bibitem [{\citenamefont {Valdez}\ \emph
  {et~al.}(2012{\natexlab{a}})\citenamefont {Valdez}, \citenamefont {Macri},\
  and\ \citenamefont {Braunstein}}]{Val_12}%
  \BibitemOpen
  \bibfield  {author} {\bibinfo {author} {\bibfnamefont {L.~D.}\ \bibnamefont
  {Valdez}}, \bibinfo {author} {\bibfnamefont {P.~A.}\ \bibnamefont {Macri}}, \
  and\ \bibinfo {author} {\bibfnamefont {L.~A.}\ \bibnamefont {Braunstein}},\
  }\href@noop {} {\bibfield  {journal} {\bibinfo  {journal} {PLOS ONE}\
  }\textbf {\bibinfo {volume} {7}},\ \bibinfo {pages} {1} (\bibinfo {year}
  {2012}{\natexlab{a}})}\BibitemShut {NoStop}%
\bibitem [{\citenamefont {Volz}\ and\ \citenamefont {Meyers}(2007)}]{volz_07}%
  \BibitemOpen
  \bibfield  {author} {\bibinfo {author} {\bibfnamefont {E.}~\bibnamefont
  {Volz}}\ and\ \bibinfo {author} {\bibfnamefont {L.~A.}\ \bibnamefont
  {Meyers}},\ }\href@noop {} {\bibfield  {journal} {\bibinfo  {journal}
  {Proceedings of the Royal Society of London B: Biological Sciences}\ }\textbf
  {\bibinfo {volume} {274}},\ \bibinfo {pages} {2925} (\bibinfo {year}
  {2007})}\BibitemShut {NoStop}%
\bibitem [{\citenamefont {Xia}\ \emph {et~al.}(2012)\citenamefont {Xia},
  \citenamefont {Wang}, \citenamefont {Sun},\ and\ \citenamefont
  {Wang}}]{xia_12}%
  \BibitemOpen
  \bibfield  {author} {\bibinfo {author} {\bibfnamefont {C.}~\bibnamefont
  {Xia}}, \bibinfo {author} {\bibfnamefont {L.}~\bibnamefont {Wang}}, \bibinfo
  {author} {\bibfnamefont {S.}~\bibnamefont {Sun}}, \ and\ \bibinfo {author}
  {\bibfnamefont {J.}~\bibnamefont {Wang}},\ }\href@noop {} {\bibfield
  {journal} {\bibinfo  {journal} {Nonlinear Dynamics}\ }\textbf {\bibinfo
  {volume} {69}},\ \bibinfo {pages} {927} (\bibinfo {year} {2012})}\BibitemShut
  {NoStop}%
\bibitem [{\citenamefont {Buono}\ \emph {et~al.}(2014)\citenamefont {Buono},
  \citenamefont {Alvarez-Zuzek}, \citenamefont {Braunstein},\ and\
  \citenamefont {Macri}}]{Buono_14}%
  \BibitemOpen
  \bibfield  {author} {\bibinfo {author} {\bibfnamefont {C.}~\bibnamefont
  {Buono}}, \bibinfo {author} {\bibfnamefont {L.~G.}\ \bibnamefont
  {Alvarez-Zuzek}}, \bibinfo {author} {\bibfnamefont {L.~A.}\ \bibnamefont
  {Braunstein}}, \ and\ \bibinfo {author} {\bibfnamefont {P.~A.}\ \bibnamefont
  {Macri}},\ }\href@noop {} {\bibfield  {journal} {\bibinfo  {journal} {PLOS
  ONE}\ }\textbf {\bibinfo {volume} {9}},\ \bibinfo {pages} {e92200} (\bibinfo
  {year} {2014})}\BibitemShut {NoStop}%
\bibitem [{\citenamefont {Buono}\ and\ \citenamefont
  {Braunstein}(2015)}]{Buono_15}%
  \BibitemOpen
  \bibfield  {author} {\bibinfo {author} {\bibfnamefont {C.}~\bibnamefont
  {Buono}}\ and\ \bibinfo {author} {\bibfnamefont {L.~A.}\ \bibnamefont
  {Braunstein}},\ }\href@noop {} {\bibfield  {journal} {\bibinfo  {journal}
  {EPL (Europhysics Letters)}\ }\textbf {\bibinfo {volume} {109}},\ \bibinfo
  {pages} {26001} (\bibinfo {year} {2015})}\BibitemShut {NoStop}%
\bibitem [{\citenamefont {Alvarez-Zuzek}\ \emph
  {et~al.}(2015{\natexlab{a}})\citenamefont {Alvarez-Zuzek}, \citenamefont
  {Stanley},\ and\ \citenamefont {Braunstein}}]{Alvarez-zuzek_15}%
  \BibitemOpen
  \bibfield  {author} {\bibinfo {author} {\bibfnamefont {L.~G.}\ \bibnamefont
  {Alvarez-Zuzek}}, \bibinfo {author} {\bibfnamefont {H.~E.}\ \bibnamefont
  {Stanley}}, \ and\ \bibinfo {author} {\bibfnamefont {L.~A.}\ \bibnamefont
  {Braunstein}},\ }\href@noop {} {\bibfield  {journal} {\bibinfo  {journal}
  {Scientific Reports}\ }\textbf {\bibinfo {volume} {5}},\ \bibinfo {pages}
  {12151} (\bibinfo {year} {2015}{\natexlab{a}})}\BibitemShut {NoStop}%
\bibitem [{\citenamefont {Lagorio}\ \emph {et~al.}(2011)\citenamefont
  {Lagorio}, \citenamefont {Dickison}, \citenamefont {Vazquez}, \citenamefont
  {Braunstein}, \citenamefont {Macri}, \citenamefont {Migueles}, \citenamefont
  {Havlin},\ and\ \citenamefont {Stanley}}]{Lag_01}%
  \BibitemOpen
  \bibfield  {author} {\bibinfo {author} {\bibfnamefont {C.}~\bibnamefont
  {Lagorio}}, \bibinfo {author} {\bibfnamefont {M.}~\bibnamefont {Dickison}},
  \bibinfo {author} {\bibfnamefont {F.}~\bibnamefont {Vazquez}}, \bibinfo
  {author} {\bibfnamefont {L.~A.}\ \bibnamefont {Braunstein}}, \bibinfo
  {author} {\bibfnamefont {P.~A.}\ \bibnamefont {Macri}}, \bibinfo {author}
  {\bibfnamefont {M.~V.}\ \bibnamefont {Migueles}}, \bibinfo {author}
  {\bibfnamefont {S.}~\bibnamefont {Havlin}}, \ and\ \bibinfo {author}
  {\bibfnamefont {H.~E.}\ \bibnamefont {Stanley}},\ }\href@noop {} {\bibfield
  {journal} {\bibinfo  {journal} {Phys. Rev. E}\ }\textbf {\bibinfo {volume}
  {83}},\ \bibinfo {pages} {026102} (\bibinfo {year} {2011})}\BibitemShut
  {NoStop}%
\bibitem [{\citenamefont {Alvarez-Zuzek}\ \emph
  {et~al.}(2015{\natexlab{b}})\citenamefont {Alvarez-Zuzek}, \citenamefont
  {Buono},\ and\ \citenamefont {Braunstein}}]{alvarez2015}%
  \BibitemOpen
  \bibfield  {author} {\bibinfo {author} {\bibfnamefont {L.~G.}\ \bibnamefont
  {Alvarez-Zuzek}}, \bibinfo {author} {\bibfnamefont {C.}~\bibnamefont
  {Buono}}, \ and\ \bibinfo {author} {\bibfnamefont {L.~A.}\ \bibnamefont
  {Braunstein}},\ }in\ \href@noop {} {\emph {\bibinfo {booktitle} {Journal of
  Physics Conference Series}}},\ Vol.\ \bibinfo {volume} {640}\ (\bibinfo
  {year} {2015})\ p.\ \bibinfo {pages} {012007}\BibitemShut {NoStop}%
\bibitem [{\citenamefont {Valdez}\ \emph
  {et~al.}(2012{\natexlab{b}})\citenamefont {Valdez}, \citenamefont {Macri},\
  and\ \citenamefont {Braunstein}}]{valdez2012}%
  \BibitemOpen
  \bibfield  {author} {\bibinfo {author} {\bibfnamefont {L.~D.}\ \bibnamefont
  {Valdez}}, \bibinfo {author} {\bibfnamefont {P.~A.}\ \bibnamefont {Macri}}, \
  and\ \bibinfo {author} {\bibfnamefont {L.~A.}\ \bibnamefont {Braunstein}},\
  }\href@noop {} {\bibfield  {journal} {\bibinfo  {journal} {Phys. Rev. E}\
  }\textbf {\bibinfo {volume} {85}},\ \bibinfo {pages} {036108} (\bibinfo
  {year} {2012}{\natexlab{b}})}\BibitemShut {NoStop}%
\bibitem [{\citenamefont {Valdez}\ \emph {et~al.}(2014)\citenamefont {Valdez},
  \citenamefont {{Arag{\~a}o R{\^e}go}}, \citenamefont {Stanley},\ and\
  \citenamefont {Braunstein}}]{Valdez_Ebola}%
  \BibitemOpen
  \bibfield  {author} {\bibinfo {author} {\bibfnamefont {L.~D.}\ \bibnamefont
  {Valdez}}, \bibinfo {author} {\bibfnamefont {H.~H.}\ \bibnamefont
  {{Arag{\~a}o R{\^e}go}}}, \bibinfo {author} {\bibfnamefont {H.~E.}\
  \bibnamefont {Stanley}}, \ and\ \bibinfo {author} {\bibfnamefont {L.~A.}\
  \bibnamefont {Braunstein}},\ }\href@noop {} {\bibfield  {journal} {\bibinfo
  {journal} {Sci. Rep.}\ }\textbf {\bibinfo {volume} {5}},\ \bibinfo {pages}
  {12172} (\bibinfo {year} {2014})}\BibitemShut {NoStop}%
\bibitem [{\citenamefont {Gomes}\ \emph {et~al.}(2014)\citenamefont {Gomes},
  \citenamefont {{Pastore y Piontti}}, \citenamefont {Rossi}, \citenamefont
  {Chao}, \citenamefont {Longini}, \citenamefont {Halloran},\ and\
  \citenamefont {Vespignani}}]{Gomes_14}%
  \BibitemOpen
  \bibfield  {author} {\bibinfo {author} {\bibfnamefont {M.~F.~C.}\
  \bibnamefont {Gomes}}, \bibinfo {author} {\bibfnamefont {A.}~\bibnamefont
  {{Pastore y Piontti}}}, \bibinfo {author} {\bibfnamefont {L.}~\bibnamefont
  {Rossi}}, \bibinfo {author} {\bibfnamefont {D.}~\bibnamefont {Chao}},
  \bibinfo {author} {\bibfnamefont {I.}~\bibnamefont {Longini}}, \bibinfo
  {author} {\bibfnamefont {M.~E.}\ \bibnamefont {Halloran}}, \ and\ \bibinfo
  {author} {\bibfnamefont {A.}~\bibnamefont {Vespignani}},\ }\href@noop {}
  {\bibfield  {journal} {\bibinfo  {journal} {PLOS Current Outbreaks}\ }
  (\bibinfo {year} {2014})}\BibitemShut {NoStop}%
\bibitem [{\citenamefont {Gsell}\ \emph {et~al.}(2017)\citenamefont {Gsell},
  \citenamefont {Camacho}, \citenamefont {J~Kucharski}, \citenamefont
  {H~Watson}, \citenamefont {Bagayoko}, \citenamefont {Danmadji~Nadlaou} \emph
  {et~al.}}]{lancet_ebola}%
  \BibitemOpen
  \bibfield  {author} {\bibinfo {author} {\bibfnamefont {P.~S.}\ \bibnamefont
  {Gsell}}, \bibinfo {author} {\bibfnamefont {A.}~\bibnamefont {Camacho}},
  \bibinfo {author} {\bibfnamefont {A.}~\bibnamefont {J~Kucharski}}, \bibinfo
  {author} {\bibfnamefont {C.}~\bibnamefont {H~Watson}}, \bibinfo {author}
  {\bibfnamefont {A.}~\bibnamefont {Bagayoko}}, \bibinfo {author}
  {\bibfnamefont {S.~V.}\ \bibnamefont {Danmadji~Nadlaou}},  \emph {et~al.},\
  }\href@noop {} {\bibfield  {journal} {\bibinfo  {journal} {The Lancet
  Infectious Diseases}\ }\textbf {\bibinfo {volume} {17}},\ \bibinfo {pages}
  {1276} (\bibinfo {year} {2017})}\BibitemShut {NoStop}%
\bibitem [{\citenamefont {Fenner}\ \emph {et~al.}(1988)\citenamefont {Fenner},
  \citenamefont {Henderson}, \citenamefont {Arita}, \citenamefont {Jezek},\
  and\ \citenamefont {Ladnyi}}]{Smallpox}%
  \BibitemOpen
  \bibfield  {author} {\bibinfo {author} {\bibfnamefont {F.}~\bibnamefont
  {Fenner}}, \bibinfo {author} {\bibfnamefont {D.~A.}\ \bibnamefont
  {Henderson}}, \bibinfo {author} {\bibfnamefont {I.}~\bibnamefont {Arita}},
  \bibinfo {author} {\bibfnamefont {Z.}~\bibnamefont {Jezek}}, \ and\ \bibinfo
  {author} {\bibfnamefont {I.~D.}\ \bibnamefont {Ladnyi}},\ }\href@noop {}
  {\emph {\bibinfo {title} {Smallpox and its eradication}}}\ (\bibinfo
  {publisher} {World Health Organization},\ \bibinfo {year} {1988})\BibitemShut
  {NoStop}%
\bibitem [{\citenamefont {Lokhov}\ and\ \citenamefont {Saad}(2017)}]{Lokhov}%
  \BibitemOpen
  \bibfield  {author} {\bibinfo {author} {\bibfnamefont {A.}~\bibnamefont
  {Lokhov}, \bibfnamefont {Y.}}\ and\ \bibinfo {author} {\bibfnamefont
  {D.}~\bibnamefont {Saad}},\ }\href@noop {} {\bibfield  {journal} {\bibinfo
  {journal} {Proceedings of the National Academy of Sciences}\ }\textbf
  {\bibinfo {volume} {14}},\ \bibinfo {pages} {8138} (\bibinfo {year}
  {2017})}\BibitemShut {NoStop}%
\bibitem [{\citenamefont {Tragler}\ \emph {et~al.}(2001)\citenamefont
  {Tragler}, \citenamefont {Caulkins},\ and\ \citenamefont
  {Feichtinger}}]{Drug}%
  \BibitemOpen
  \bibfield  {author} {\bibinfo {author} {\bibfnamefont {G.}~\bibnamefont
  {Tragler}}, \bibinfo {author} {\bibfnamefont {J.~P.}\ \bibnamefont
  {Caulkins}}, \ and\ \bibinfo {author} {\bibfnamefont {G.}~\bibnamefont
  {Feichtinger}},\ }\href@noop {} {\bibfield  {journal} {\bibinfo  {journal}
  {Operations Research}\ }\textbf {\bibinfo {volume} {49}},\ \bibinfo {pages}
  {325} (\bibinfo {year} {2001})}\BibitemShut {NoStop}%
\bibitem [{\citenamefont {Miller}\ \emph {et~al.}(2012)\citenamefont {Miller},
  \citenamefont {Slim},\ and\ \citenamefont {Volz}}]{miller_12}%
  \BibitemOpen
  \bibfield  {author} {\bibinfo {author} {\bibfnamefont {J.~C.}\ \bibnamefont
  {Miller}}, \bibinfo {author} {\bibfnamefont {A.~C.}\ \bibnamefont {Slim}}, \
  and\ \bibinfo {author} {\bibfnamefont {E.~M.}\ \bibnamefont {Volz}},\
  }\href@noop {} {\bibfield  {journal} {\bibinfo  {journal} {Journal of the
  Royal Society Interface}\ }\textbf {\bibinfo {volume} {9}},\ \bibinfo {pages}
  {890} (\bibinfo {year} {2012})}\BibitemShut {NoStop}%
\bibitem [{\citenamefont {B\"otcher}\ \emph {et~al.}(2015)\citenamefont
  {B\"otcher}, \citenamefont {Woolley-Meza}, \citenamefont {Ara\'ujo},
  \citenamefont {Herrmann},\ and\ \citenamefont {Helbing}}]{Bottcher_15}%
  \BibitemOpen
  \bibfield  {author} {\bibinfo {author} {\bibfnamefont {L.}~\bibnamefont
  {B\"otcher}}, \bibinfo {author} {\bibfnamefont {O.}~\bibnamefont
  {Woolley-Meza}}, \bibinfo {author} {\bibfnamefont {N.~A.}\ \bibnamefont
  {Ara\'ujo}}, \bibinfo {author} {\bibfnamefont {H.~J.}\ \bibnamefont
  {Herrmann}}, \ and\ \bibinfo {author} {\bibfnamefont {D.}~\bibnamefont
  {Helbing}},\ }\href@noop {} {\bibfield  {journal} {\bibinfo  {journal}
  {Scientific Reports}\ }\textbf {\bibinfo {volume} {5}},\ \bibinfo {pages}
  {16571} (\bibinfo {year} {2015})}\BibitemShut {NoStop}%
\bibitem [{\citenamefont {Chen}\ \emph {et~al.}(2016)\citenamefont {Chen},
  \citenamefont {Zhou}, \citenamefont {Feng}, \citenamefont {Liang},
  \citenamefont {Liljeros}, \citenamefont {Havlin},\ and\ \citenamefont
  {Hu}}]{Chen2}%
  \BibitemOpen
  \bibfield  {author} {\bibinfo {author} {\bibfnamefont {X.}~\bibnamefont
  {Chen}}, \bibinfo {author} {\bibfnamefont {T.}~\bibnamefont {Zhou}}, \bibinfo
  {author} {\bibfnamefont {L.}~\bibnamefont {Feng}}, \bibinfo {author}
  {\bibfnamefont {J.}~\bibnamefont {Liang}}, \bibinfo {author} {\bibfnamefont
  {F.}~\bibnamefont {Liljeros}}, \bibinfo {author} {\bibfnamefont
  {S.}~\bibnamefont {Havlin}}, \ and\ \bibinfo {author} {\bibfnamefont
  {Y.}~\bibnamefont {Hu}},\ }\href@noop {} {\bibfield  {journal} {\bibinfo
  {journal} {ArXiv e-prints}\ } (\bibinfo {year} {2016})},\ \Eprint
  {http://arxiv.org/abs/1611.00212} {arXiv:1611.00212 [physics.soc-ph]}
  \BibitemShut {NoStop}%
\bibitem [{\citenamefont {Chen}\ \emph {et~al.}(2018)\citenamefont {Chen},
  \citenamefont {Wang}, \citenamefont {Tang}, \citenamefont {Cai},
  \citenamefont {Stanley},\ and\ \citenamefont {Braunstein}}]{Chen_18}%
  \BibitemOpen
  \bibfield  {author} {\bibinfo {author} {\bibfnamefont {X.}~\bibnamefont
  {Chen}}, \bibinfo {author} {\bibfnamefont {R.}~\bibnamefont {Wang}}, \bibinfo
  {author} {\bibfnamefont {M.}~\bibnamefont {Tang}}, \bibinfo {author}
  {\bibfnamefont {S.}~\bibnamefont {Cai}}, \bibinfo {author} {\bibfnamefont
  {H.~E.}\ \bibnamefont {Stanley}}, \ and\ \bibinfo {author} {\bibfnamefont
  {L.~A.}\ \bibnamefont {Braunstein}},\ }\href@noop {} {\bibfield  {journal}
  {\bibinfo  {journal} {New Journal of Physics}\ }\textbf {\bibinfo {volume}
  {20}},\ \bibinfo {pages} {013007} (\bibinfo {year} {2018})}\BibitemShut
  {NoStop}%
\bibitem [{\citenamefont {{Alvarez-Zuzek}}\ \emph {et~al.}(2018)\citenamefont
  {{Alvarez-Zuzek}}, \citenamefont {{Di Muro}}, \citenamefont {{Havlin}},\ and\
  \citenamefont {{Braunstein}}}]{Alvarez_2018}%
  \BibitemOpen
  \bibfield  {author} {\bibinfo {author} {\bibfnamefont {L.~G.}\ \bibnamefont
  {{Alvarez-Zuzek}}}, \bibinfo {author} {\bibfnamefont {M.~A.}\ \bibnamefont
  {{Di Muro}}}, \bibinfo {author} {\bibfnamefont {S.}~\bibnamefont {{Havlin}}},
  \ and\ \bibinfo {author} {\bibfnamefont {L.~A.}\ \bibnamefont
  {{Braunstein}}},\ }\href@noop {} {\bibfield  {journal} {\bibinfo  {journal}
  {ArXiv e-prints}\ } (\bibinfo {year} {2018})},\ \Eprint
  {http://arxiv.org/abs/1804.10593} {arXiv:1804.10593 [physics.soc-ph]}
  \BibitemShut {NoStop}%
\bibitem [{\citenamefont {Newman}\ \emph {et~al.}(2001)\citenamefont {Newman},
  \citenamefont {Strogatz},\ and\ \citenamefont {Watts}}]{New_03}%
  \BibitemOpen
  \bibfield  {author} {\bibinfo {author} {\bibfnamefont {M.~E.~J.}\
  \bibnamefont {Newman}}, \bibinfo {author} {\bibfnamefont {S.~H.}\
  \bibnamefont {Strogatz}}, \ and\ \bibinfo {author} {\bibfnamefont {D.~J.}\
  \bibnamefont {Watts}},\ }\href@noop {} {\bibfield  {journal} {\bibinfo
  {journal} {Phys. Rev. E}\ }\textbf {\bibinfo {volume} {64}},\ \bibinfo
  {pages} {026118} (\bibinfo {year} {2001})}\BibitemShut {NoStop}%
\bibitem [{\citenamefont {Newman}(2003)}]{New_01}%
  \BibitemOpen
  \bibfield  {author} {\bibinfo {author} {\bibfnamefont {M.~E.~J.}\
  \bibnamefont {Newman}},\ }\href@noop {} {\bibfield  {journal} {\bibinfo
  {journal} {Phys. Rev. E}\ }\textbf {\bibinfo {volume} {67}},\ \bibinfo
  {pages} {026126} (\bibinfo {year} {2003})}\BibitemShut {NoStop}%
\bibitem [{\citenamefont {Molloy}\ and\ \citenamefont {Reed}(1995)}]{Mol_01}%
  \BibitemOpen
  \bibfield  {author} {\bibinfo {author} {\bibfnamefont {M.}~\bibnamefont
  {Molloy}}\ and\ \bibinfo {author} {\bibfnamefont {B.}~\bibnamefont {Reed}},\
  }\href@noop {} {\bibfield  {journal} {\bibinfo  {journal} {Random Structures
  and Algorithms}\ }\textbf {\bibinfo {volume} {6}},\ \bibinfo {pages} {161}
  (\bibinfo {year} {1995})}\BibitemShut {NoStop}%
\bibitem [{\citenamefont {Cohen}\ \emph {et~al.}(2000)\citenamefont {Cohen},
  \citenamefont {Erez}, \citenamefont {{ben-Avraham}},\ and\ \citenamefont
  {Havlin}}]{Coh_01}%
  \BibitemOpen
  \bibfield  {author} {\bibinfo {author} {\bibfnamefont {R.}~\bibnamefont
  {Cohen}}, \bibinfo {author} {\bibfnamefont {K.}~\bibnamefont {Erez}},
  \bibinfo {author} {\bibfnamefont {D.}~\bibnamefont {{ben-Avraham}}}, \ and\
  \bibinfo {author} {\bibfnamefont {S.}~\bibnamefont {Havlin}},\ }\href@noop {}
  {\bibfield  {journal} {\bibinfo  {journal} {Phys. Rev. Lett.}\ }\textbf
  {\bibinfo {volume} {85}},\ \bibinfo {pages} {4626} (\bibinfo {year}
  {2000})}\BibitemShut {NoStop}%
\bibitem [{\citenamefont {Parshani}\ \emph {et~al.}(2011)\citenamefont
  {Parshani}, \citenamefont {Buldyrev},\ and\ \citenamefont {Havlin}}]{Par_02}%
  \BibitemOpen
  \bibfield  {author} {\bibinfo {author} {\bibfnamefont {R.}~\bibnamefont
  {Parshani}}, \bibinfo {author} {\bibfnamefont {S.~V.}\ \bibnamefont
  {Buldyrev}}, \ and\ \bibinfo {author} {\bibfnamefont {S.}~\bibnamefont
  {Havlin}},\ }\href@noop {} {\bibfield  {journal} {\bibinfo  {journal} {Proc.
  Natl. Acad. Sci.}\ }\textbf {\bibinfo {volume} {108}},\ \bibinfo {pages}
  {1007} (\bibinfo {year} {2011})}\BibitemShut {NoStop}%
\bibitem [{\citenamefont {Zhou}\ \emph {et~al.}(2014)\citenamefont {Zhou},
  \citenamefont {Bashan}, \citenamefont {Cohen}, \citenamefont {Berezin},
  \citenamefont {Shnerb},\ and\ \citenamefont {Havlin}}]{Zhou14}%
  \BibitemOpen
  \bibfield  {author} {\bibinfo {author} {\bibfnamefont {D.}~\bibnamefont
  {Zhou}}, \bibinfo {author} {\bibfnamefont {A.}~\bibnamefont {Bashan}},
  \bibinfo {author} {\bibfnamefont {R.}~\bibnamefont {Cohen}}, \bibinfo
  {author} {\bibfnamefont {Y.}~\bibnamefont {Berezin}}, \bibinfo {author}
  {\bibfnamefont {N.}~\bibnamefont {Shnerb}}, \ and\ \bibinfo {author}
  {\bibfnamefont {S.}~\bibnamefont {Havlin}},\ }\href@noop {} {\bibfield
  {journal} {\bibinfo  {journal} {Phys. Rev. E}\ }\textbf {\bibinfo {volume}
  {90}},\ \bibinfo {pages} {012803} (\bibinfo {year} {2014})}\BibitemShut
  {NoStop}%
\bibitem [{\citenamefont {Callaway}\ \emph {et~al.}(2000)\citenamefont
  {Callaway}, \citenamefont {Newman}, \citenamefont {Strogatz},\ and\
  \citenamefont {Watts}}]{Dun_01}%
  \BibitemOpen
  \bibfield  {author} {\bibinfo {author} {\bibfnamefont {D.~S.}\ \bibnamefont
  {Callaway}}, \bibinfo {author} {\bibfnamefont {M.~E.~J.}\ \bibnamefont
  {Newman}}, \bibinfo {author} {\bibfnamefont {S.~H.}\ \bibnamefont
  {Strogatz}}, \ and\ \bibinfo {author} {\bibfnamefont {D.~J.}\ \bibnamefont
  {Watts}},\ }\href@noop {} {\bibfield  {journal} {\bibinfo  {journal} {Phys.
  Rev. Lett.}\ }\textbf {\bibinfo {volume} {85}},\ \bibinfo {pages} {5468}
  (\bibinfo {year} {2000})}\BibitemShut {NoStop}%
\bibitem [{\citenamefont {Vel{\'a}squez-Rojas}\ and\ \citenamefont
  {Vazquez}(2017)}]{velasquez17}%
  \BibitemOpen
  \bibfield  {author} {\bibinfo {author} {\bibfnamefont {F.}~\bibnamefont
  {Vel{\'a}squez-Rojas}}\ and\ \bibinfo {author} {\bibfnamefont
  {F.}~\bibnamefont {Vazquez}},\ }\href@noop {} {\bibfield  {journal} {\bibinfo
   {journal} {Physical Review E}\ }\textbf {\bibinfo {volume} {95}},\ \bibinfo
  {pages} {052315} (\bibinfo {year} {2017})}\BibitemShut {NoStop}%
\bibitem [{\citenamefont {G{\'o}mez-Garde{\~n}es}\ \emph
  {et~al.}(2016)\citenamefont {G{\'o}mez-Garde{\~n}es}, \citenamefont {Lotero},
  \citenamefont {Taraskin},\ and\ \citenamefont {P{\'e}rez-Reche}}]{gomez16}%
  \BibitemOpen
  \bibfield  {author} {\bibinfo {author} {\bibfnamefont {J.}~\bibnamefont
  {G{\'o}mez-Garde{\~n}es}}, \bibinfo {author} {\bibfnamefont {L.}~\bibnamefont
  {Lotero}}, \bibinfo {author} {\bibfnamefont {S.}~\bibnamefont {Taraskin}}, \
  and\ \bibinfo {author} {\bibfnamefont {F.}~\bibnamefont {P{\'e}rez-Reche}},\
  }\href@noop {} {\bibfield  {journal} {\bibinfo  {journal} {Scientific
  reports}\ }\textbf {\bibinfo {volume} {6}},\ \bibinfo {pages} {19767}
  (\bibinfo {year} {2016})}\BibitemShut {NoStop}%
\end{thebibliography}

%

\end{document}